\documentclass{article}

\usepackage{arxiv}

\usepackage[utf8]{inputenc} 
\usepackage[T1]{fontenc}    
\usepackage{hyperref}       
\usepackage{url}            
\usepackage{booktabs}       
\usepackage{amsfonts}       
\usepackage{nicefrac}       
\usepackage{microtype}      
\hypersetup{
  colorlinks   = true, 
  urlcolor     = blue, 
  linkcolor    = blue, 
  citecolor   = blue 
}

\usepackage{tikz}

\usepackage[square,numbers]{natbib}
\usepackage{graphicx}
\usepackage{algorithmic}
\usepackage{graphicx}
\usepackage{textcomp}
\usepackage{xcolor}
\usepackage[colorinlistoftodos,prependcaption,textsize=tiny]{todonotes}
\usepackage{balance}
\usepackage{booktabs} %
\usepackage{easyReview}
\usepackage{gensymb}
\usepackage[utf8]{inputenc}
\usepackage[T1]{fontenc}
\usepackage{multirow}
\usepackage{float}
\usepackage{graphicx}
\usepackage{subfig}
\usepackage{makecell}
\usepackage{enumitem}
\usepackage{pgfplots}
\usepackage{amssymb, amsmath, amsthm}
\usepackage{chronology}

\usepackage{placeins}
\usepackage{adjustbox}
\usepackage[title]{appendix}
\usepackage{pifont}
\usepackage{colortbl}
\usepackage{lineno}
\usepackage{etoolbox}
\usepackage{multicol}

\usetikzlibrary{decorations.text}
\usetikzlibrary{decorations.pathreplacing}
\usetikzlibrary{decorations.pathreplacing,positioning, arrows.meta}

\pgfplotsset{compat=1.7}
\pgfplotsset{every x tick label/.append style={font=\tiny, yshift=0.5ex}}

\usetikzlibrary{shapes,arrows,arrows.meta,chains,matrix,positioning,decorations.pathmorphing,decorations.pathreplacing,fit,calc,backgrounds,hobby,bending,automata,trees}

\definecolor{carmine}{rgb}{0.59, 0.0, 0.09}
\definecolor{dartmouthgreen}{rgb}{0.05, 0.5, 0.06}

\usepackage{placeins}
\restylefloat{table}
\makeatletter
\pgfkeys{%
  /pgf/decoration/.cd,
  start color/.store in =\startcolor,
  end color/.store in   =\endcolor,
  start width/.store in =\startwidth,
  end width/.store in   =\endwidth,
  start color=black!5,
  end color=black,
}
\pgfdeclaredecoration{width and color change}{initial}{
 \state{initial}[width=0pt, next state=line, persistent precomputation={%
   \pgfmathdivide{50}{\pgfdecoratedpathlength}%
   \let\increment=\pgfmathresult%
   \def\x{0}%
 }]{}
 \state{line}[width=.5pt, persistent postcomputation={%
     \pgfmathadd@{\x}{\increment}%
     \let\x=\pgfmathresult%
   }]{%
   \pgfsetlinewidth{\x/40*0.075pt+\pgflinewidth}%
   \pgfsetarrows{-}%
   \pgfpathmoveto{\pgfpointorigin}%
   \pgfpathlineto{\pgfqpoint{.75pt}{0pt}}%
   \pgfsetstrokecolor{\endcolor!\x!\startcolor}%
   \pgfusepath{stroke}%
 }
 \state{final}{%
   \pgfsetlinewidth{\pgflinewidth}%
   \pgfpathmoveto{\pgfpointorigin}%
   \color{\endcolor!\x!\startcolor}%
   \pgfusepath{stroke}%
 }
}
\makeatother

\theoremstyle{plain}
\newtheorem*{definition}{Definition}
\newtheorem*{example}{Example}

\usepackage{array}
\newcommand\MyBox[2]{
  \fbox{\lower0.75cm
    \vbox to 1.7cm{\vfil
      \hbox to 1.7cm{\hfil\parbox{1.4cm}{#1\\#2}\hfil}
      \vfil}%
  }%
}

\newcommand\OkBox[1]{
\setlength\fboxrule{0pt}
  \fbox{\lower0.75cm
    \vbox to 1.2cm{\vfil
      \hbox to 5cm{\hfil\parbox{4.5cm}{#1}\hfil}
      \vfil}%
  }%
}

\newcommand\MetBox[1]{
\setlength\fboxrule{0pt}
  \fbox{\lower0.75cm
    \vbox to 1.2cm{\vfil
      \hbox to 2cm{\hfil\parbox{2cm}{#1}\hfil}
      \vfil}%
  }%
}

\newcommand\ValBox[1]{
\setlength\fboxrule{0pt}
  \fbox{\lower0.75cm
    \vbox to 1.2cm{\vfil
      \hbox to 2.2cm{\hfil\parbox{2.2cm}{#1}\hfil}
      \vfil}%
  }%
}

\tikzset{My Arrow Style/.style={single arrow, fill=red!50, anchor=base, align=center,text width=2.8cm}}

\title{Impact of Weather Factors on Migration Intention using Machine Learning Algorithms}

\begin{multicols}{3}
\author{
  John Aoga\\
  Science For Engineers Doctoral School\\ 
  University of Abomey-Calavi\\
  Abomey-Calavi, Bénin\\
  \texttt{\small  johnaoga@gmail.com} \\
   \And
  Juhee Bae \\
  School of Informatics\\
  University of Skövde\\
  Skövde, Sweden \\
  \texttt{\small  juhee.bae@his.se} \\
   \AND
  Stefanija Veljanoska \\
   CNRS, CREM-UMR6211 \\ 
   Université de Rennes 1\\
   Rennes, France   \\
   \texttt{\small  stefanija.veljanoska} \\
   \texttt{\small  @univ-rennes1.fr} \\
   \And 
   Siegfried Nijssen \\
   ICTEAM \\
   Université catholique de Louvain\\
   Louvain-la-Neuve, Belgium\\
   \texttt{\small  siegfried.nijssen@uclouvain.be} \\
   \And
   Pierre Schaus \\
   ICTEAM \\
   Université catholique de Louvain\\
   Louvain-la-Neuve, Belgium\\
   \texttt{\small pierre.schaus@uclouvain.be} \\
}
\end{multicols}

\begin{document}
\maketitle

\begin{abstract}
A growing attention in the empirical literature has been paid to the incidence of climate shocks and change in migration decisions. Previous literature leads to different results and uses a multitude of traditional empirical approaches.

This paper proposes a tree-based Machine Learning (ML) approach to analyze the role of the weather shocks towards an individual's intention to migrate in the six agriculture-dependent-economy countries such as Burkina Faso, Ivory Coast, Mali, Mauritania, Niger, and Senegal. We perform several tree-based algorithms (e.g., XGB, Random Forest) using the train-validation-test workflow to build robust and noise-resistant approaches. Then we determine the important features showing in which direction they are influencing the migration intention. This ML-based estimation accounts for features such as weather shocks captured by the Standardized Precipitation-Evapotranspiration Index (SPEI) for different timescales and various socioeconomic features/covariates.

We find that (i) weather features improve the prediction performance although socioeconomic characteristics have more influence on migration intentions, (ii) country-specific model is necessary, and (iii) international move is influenced more by the longer timescales of SPEIs while general move (which includes internal move) by that of shorter timescales.
\end{abstract}


\section{Introduction}

Climate is changing, and its implication for human mobility is at the core of the scientific and political agenda. The profound relationship between migration and the environment is not an unknown phenomenon, but the emergence and acceleration of climate change introduce more complexity to this relationship. The literature bringing together migration and climate change has significantly increased in the past ten years (\citet{millock2015migration,berlemann2017climate, beine2018meta, cattaneo2019human}). This literature benefited primarily from greater availability and quality of climate and mobility indicators. Their main goal is to study the extent to which climate events initiated or even forced individuals to move. Even though the research goal seems straightforward, the findings do not reach a consensus. 

The heterogeneity of results is because of the use of different measurements, methodological approaches, and different contexts (\citet{beine2018meta}). First, findings differ in terms of the (i) direction of the impact whether climate acts as a pull or a push factor for migration\footnote{\citet{Black2013} distinguish migration, displacement, and immobility and \citet{beine2018meta} refer to `trapped population'.}, (ii) the strength of the relationship, and (iii) that this relationship is conditional on other features. Second, the different methodological approaches and ways of measuring climate shocks and migration could explain such divergence of existent evidence. Third, the findings are context specific. For example, existing evidence shows that the climate-migration nexus is common in developing societies with the rain-fed agricultural sector that occupies a vital place in the overall economy. 

The primary goal of this article is to bring new insights to this literature by adopting Machine Learning (ML) techniques and a multitude of climate and mobility measurements. Following~\citet{wthrShock2020}, we focus on the West African region, namely, Burkina Faso, Ivory Coast, Mali, Mauritania, Niger, and Senegal. We build upon the data from the Gallup World Poll (GWP) surveys~\citep{gallup2015worldwide} and the high-resolution gridded dataset by the Climatic Research Unit of the University of East Anglia~\citep{harris2020version} for constructing the climate indicators.

Unlike traditional methods used by social scientists that specify the relationships between variables, machine learning algorithms are emerging technologies that can learn data without the explicit specification of relationships.
Thus, one benefit is the lesser manual function operations compared to the methods used in econometric studies, reducing the possible bias introduced by the expertise of the modeler. Instead, ML models base on the dataset and can discover more complex relationships or patterns between more variables during the learning phase. 
Therefore, ML models are considered more as a ``black box'' approach.
This higher \textit{capacity} comes with two major drawbacks.
The first one is the risk of overfitting the model to the data.
Therefore, ML methodology generally splits the dataset of observations into the so-called \textit{training set} and \textit{test set}. The first one is used to fit the model, and the second one to evaluate the performances of the learned models.
The second drawback of ML approaches is the less interpretable models that they generate. While a linear model can be easily interpreted by looking at the weights of the coefficients of each covariate, some machine learning models can involve thousands or even millions of parameters combined with a complex mathematical or logical formula to take each decision. Some methods have been imagined helping the user interpret ML models, but in general, the behavior of ML models remains arguably less interpretable than linear methods.

Among the large variety of machine learning approaches, we have chosen in this work to use (ensemble) tree-based classifiers -- decision tree (DT), random forest (RF), and eXtreme Gradient Boosting (XGB) -- for the following reasons.
The reproducibility of our result/approach is important and these methods are available today in most of the off-the-shelf data-science tools or library.
The methods existed for nearly two decades and have obtained excellent performances on many machine learning problems. They are still considered as the method to try first in machine learning competitions.
These methods do not require too large datasets to get good results, and the training cost is quite low.

Our contributions are as follows.

\begin{itemize}
   \item We approach the migration-climate nexus using tree-based methods and demonstrate through this paper the interest to use ML. 
    \item We provide evidence on how climate influences migration intentions.
    \item We propose an ML workflow to the community of social sciences on how to use machine learning techniques.
\end{itemize}

Section \ref{sec:problem} introduces the problem and lists our questions based on our motivation for using machine learning methods to predict migration intentions. Next, Section \ref{sec:ml} (and Appendix \ref{apx:ml}) describes our methodology and provides an overview of the machine learning (ML) approaches (e.g., decision tree (DT), random forest (RF), and eXtreme Gradient Boosting (XGB)) we used in this paper. In addition, Section \ref{sec:data} describes the dataset with details. Then, Section \ref{sec:results} describes our experiments and answers to the research questions followed by Section \ref{sec:discussion} which elaborates on our findings and discussions. We finish our paper in Section \ref{sec:conclusion}.

\section{Conceptual framework}
\label{sec:problem}

\subsection{Formalizing the climate-migration relation}
\label{sec:problem1}
The idea that weather may affect economic outcomes such as economic growth, agricultural output, migration, among others, is not a recent idea \citep{dell2014we}. 
Establishing such a relationship is a long-standing challenge. 
It is challenging to separate the effect of climate from other influences (non-climatic), which are potentially correlated with it.

The climate-migration nexus is typically represented as identifying an unknown functional relationship, $f$:
\begin{equation}\label{eqn:basicmodel}
    y_{irkt} = f(C_{rkt}, X_{irkt}).
\end{equation}
This function links climate variables ($C$) available for a set of regions $r$ in a country $k$ in year $t$ and other potentially explanatory variables ($X$) available at individual level $i$ to migration outcomes ($y$). 

In this paper, $y_{irkt}$ represents \textit{migration willingness to emigrate} either locally or abroad. It is a binary variable that equals one (or `yes, I plan to migrate') when an individual is planning to move or otherwise a zero (or `no'). It is to be observed that this variable captures potential migration or a migration plan, but not an official migration flow. In fact, in general, migration intention is used as an indicator of potential future migration because of the lack of trustworthy information of the actual migration flow \citep{Tjaden2019}. \citet{Tjaden2019} validates the usefulness of emigration intentions data, especially when actual migration flow information is not available.

$C_{rkt}$ represents the weather shocks based on the multiscalar drought index computed from weather data called SPEI (Standardized Precipitation-Evapotranspiration Index) provided at the regional level (\citet{VicenteSerrano2010}). SPEI is an improved drought index that accounts for atmospheric water conditions that are affected by temperature, wind, and humidity. SPEI normalizes accumulated climatic water balance (a difference between precipitation and potential evapotranspiration) anomalies to measure the drought severity. To compare across locations and climates, log-logistic probability distribution is used for normalization, as suggested by \citet{VicenteSerrano2010}. For more details of SPEI, we refer to Section~\ref{sec:dataWeatherSPEI}.
Other variables considered in the model represent demographic (e.g., age, gender) and socioeconomic variables (e.g., income) related to a sample of individuals living in a country.

A probit model is one of the traditional approaches used to estimate Equation \eqref{eqn:basicmodel}. Its form is: 
\begin{equation}\label{eqn:linearmodel}
    P(y_{irkt}=1\mid C_{rkt}, X_{irkt}) = \Phi(\alpha + \beta C_{rkt} + \gamma X_{irkt})
\end{equation}
where $\beta$ and $\gamma$ are the parameters that characterize the contribution and the role of the regional climate ($C_{rkt}$) and individual covariates ($X_{irkt}$) for individual migration intentions. $\Phi$($\cdot$) is the cumulative standard normal distribution function.

To understand the climate determinants of migration intentions, \citet{wthrShock2020} use a logit approach to estimate migration intention decision that depends on a utility that an individual $i$ would derive from the opting for the different migration possibilities or remaining at the origin. This utility, similar to Equation \eqref{eqn:linearmodel}, depends on regional climate and other individual's demographic and socioeconomic characteristics and considers time and regional dummies. The latter controls the possible seasonal effects in the stated intentions to migrate, the time-varying country-level determinants of these intentions, and the time-invariant spatial heterogeneity in the intentions to move. The study comprises two stages of analysis. In their first stage, \citet{wthrShock2020} perform over 300,000 regressions to select the weather factors influencing migration intentions on several samples. The study uses these selected variables at the second stage to estimate their parameters and the direction of influence towards migration intentions.

\subsection{Benefits of Decision trees over linear models}
\label{subsec:benefits}

Discovering the best linear model can become very complex as the number of input variables (covariates) grows.
In~\citet{wthrShock2020}, the authors reported over 300,000 regressions performed resulting from some handcrafted terms that are added to the model (Equation~\eqref{eqn:linearmodel}) by multiplying and/or taking the logarithm of several variables. 
Also, comparing each regression model, as done in~\citet{wthrShock2020} without a training/test set approach, might cause a bias toward more complex models that may perform poorly on unseen data.
We carry out two comparative studies using the dataset described in \citet{wthrShock2020}.
\begin{enumerate}
    \item Comparing the predictive power between two different ways of running logistic regressions: \textit{several regressions} (as described in \citet{wthrShock2020}) versus \textit{single regression} (ML's way) to examine their differences\footnote{The major difference between the logistic regression from the ML approach and the regressions used in \citet{wthrShock2020} is that the logistic regression from the ML approach runs a \textit{single regression}, including all features or covariates, while in \citet{wthrShock2020}, there are multiple runs of regressions (i.e., a run for each feature).}.
    \item Comparing the predictive power of a \textit{linear} model (logistic regression) versus a \textit{non-linear} model (decision trees) to examine the nature of the problem we have.
\end{enumerate}

\begin{figure}
    \centering
    \subfloat[Logistic regression: parameter estimation (refer) versus machine learning (LR) perspectives]
   {
    \resizebox{0.75\columnwidth}{!}{
    \includegraphics{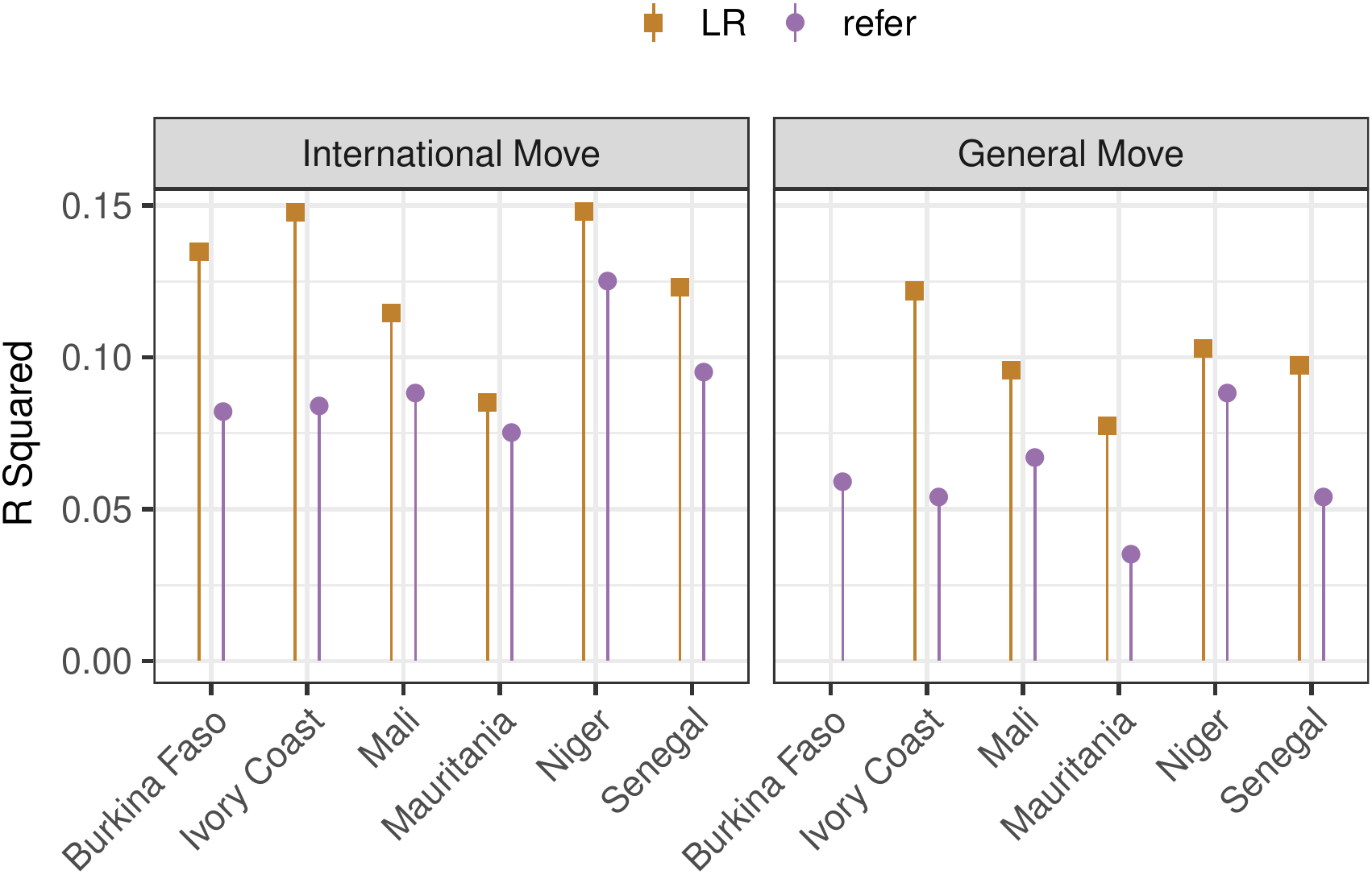}
    }
    }
    
   \subfloat[Logistic regression (LR) versus decision tree (DT) in machine learning]
   {
    \resizebox{0.95\columnwidth}{!}{
    \includegraphics{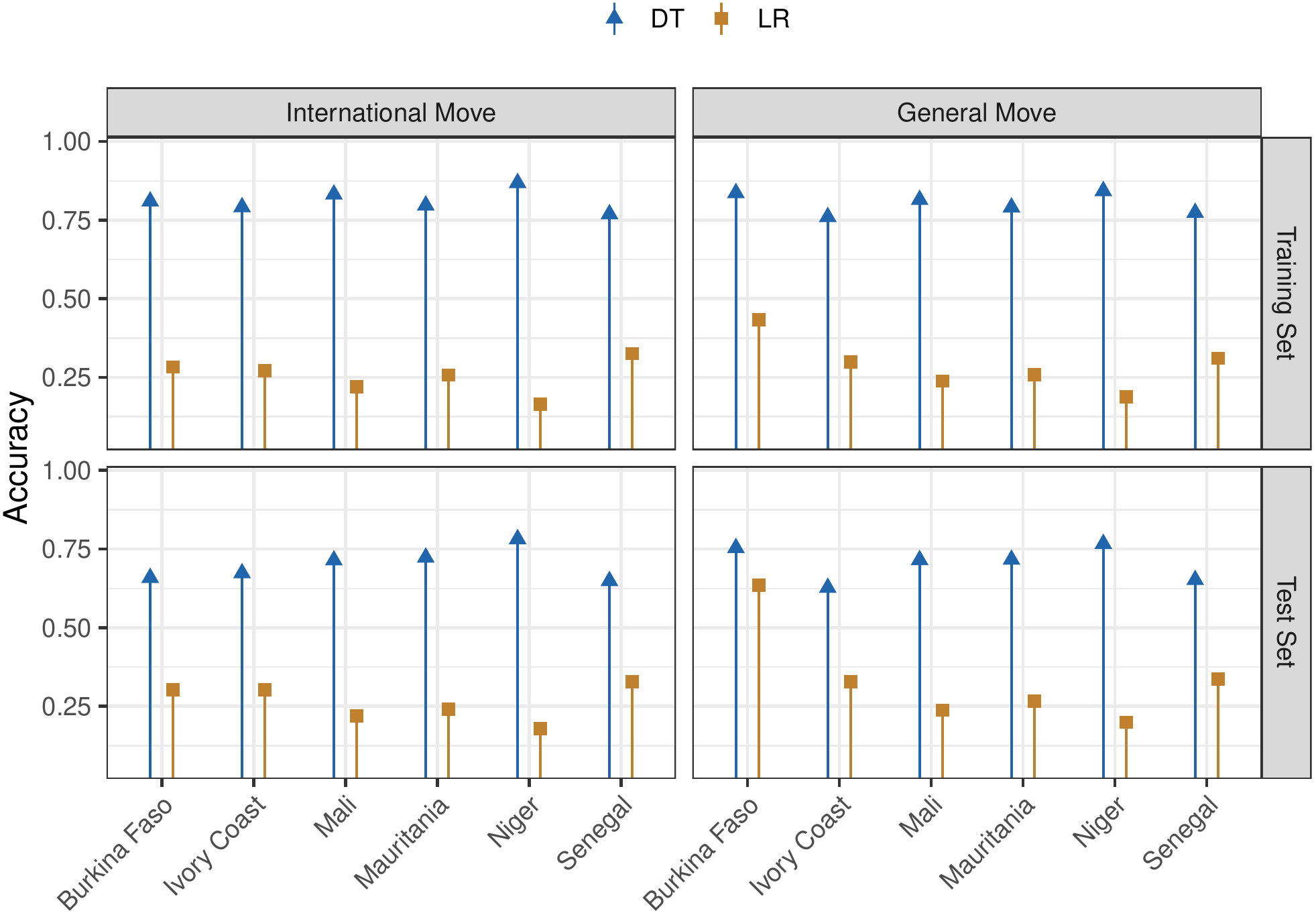}
    }
    }
    \caption{Comparing model's predictive power for each country using: \textbf{(a)} R$^{2}$ measures of logistic estimation provided in \citet{wthrShock2020} (refer) and ML's logistic regression results (LR), and \textbf{(b)} Accuracy measures of ML's logistic regression results (LR) and decision trees (DT). In (a), the ML's logistic regression model (LR) for Burkina Faso (general move) is empty because it does not converge. \label{fig:logitReg}}
\end{figure}

Figure~\ref{fig:logitReg}(a) compares the \textit{predictive power} of both regression models using R-Squared measurement\footnote{R-squared can be computed using the \textit{McFadden's} R$^{2}$ formula~\citep{mcfadden1973conditional}. \citet{wthrShock2020}, use R-squared measure implemented in \textit{STATA}~\citep{statacorp2007stata}: 
$1 - L_M/L_0,$ where $L_M$ is the log-likelihood of the model and $L_0$ is the log-likelihood of a \textit{null-model}. A null-model is a model where we learn only from the target attribute with no predictor.}.
ML's logistic regression (LR) outperforms the regressions from~\citet{wthrShock2020} (refer) in terms of the predictive power. Hence, machine learning (ML) yields higher predictive power and more interesting results. %

However, it should be noted that a low R-squared measure does not imply that the model performs worse. The models with very low R-squared can fit the data very well according to the \textit{goodness-of-fit} tests. This meets the goal of parameter estimation, whereas, in ML, the performance on unknown dataset/instances is more critical. 

In a second step, the model is trained on a part of the dataset (i.e., training set). The other part (i.e., hold-out sample or test set) is used to measure the model's predictive power of the unknown observations. There are a number of metrics to measure the predictive power of a model. Section~\ref{sec:metric:auc} provides more details of the metrics.

The accuracy value is between $0$ and $1$, same as R-squared (higher values, the better).
Figure~\ref{fig:logitReg}(b) shows the accuracy of a logistic regression model (LR) and a decision tree (DT). 
DT outperforms the LR on both training and test sets. However, accuracy is not a reliable measure when class distributions show severe skewness (\citet{Branco2016}). But it is common to face this imbalanced distribution with the real dataset (refer to the counts in Figure \ref{fig:datasetsizes}). The positive migration intentions (minority class) are more important to our analysis, but accuracy shows a lesser impact on this minor representation as depicted in Figure 1(b). It is why we show alternative metrics for an imbalanced classification problem in Sections \ref{sec:metric:auc} and \ref{sec:results}. Overall, tree-based approaches better capture the nonlinearity of our problem.

Typically, the learning workflow includes (i) selecting a proper ML method suitable to the problem taking into account several criteria: the quality of the data (e.g., reduce noise in data), the linearity of the problem, and the interpretability of the outcomes. Then, (ii) we optimize the method’s configuration to improve the overall model’s predictive power. Finally, (iii) we interpret the outcomes of the models to provide insights into the problem.

In this paper, we use tree-based ML methods to overcome the problems (e.g., scalability issues solving nonlinearity, identify driving factors) raised in previous literature when examining the relationship between migration and climate. We closely follow the data construction of \citet{wthrShock2020} to study this relationship. The difference is that the present study involves the actual SPEI values with no transformation over longer periods and uses tree-based methods to predict the migration intention and capture the non-linear relationships of the input variables.

\subsection{Toward ML approaches}
\label{sec:questions}
To understand the links between the climate, individual characteristics, and migration intentions in a more flexible methodological manner, we propose to use machine learning (ML) algorithms. It makes it possible to predict the migration intention with a larger dataset and find influencing features from an existing explainable method (Section \ref{sec:pdp}).

We first find an ML approach that shows robust prediction performances (Q1). In addition, with a large dataset including the weather shock and individual characteristics from a survey on migration intentions of six different countries, we statistically compare the prediction performance to find the impact of weather (Q2, Q5), individual characteristics (Q3, Q4), and countries (Q3, Q4). The weather shock involves various timescales of SPEI and different lengths of the lags before the interview date (over 4 years from the interview date) (Q5).

\begin{itemize}[noitemsep,topsep=0pt]
	\item Q1: Which tree-based ML algorithm(s) performs better, i.e., with a higher score?
	\item Q2: Does weather (i.e., drought) influence moving intentions? 
	\item Q3: Can we generalize a model for the six countries or need a country-specific model?
	\item Q4: Which features impact moving intentions?
	\item Q5: Does SPEI index or monthly lags of weather influence moving intentions, and if so, which SPEI(s) or lag(s) matter?
\end{itemize}

For the sake of simplicity, in the remainder of the paper, we will use $X$ to denote the dataset used by learning models without distinguishing between climate and control variables unless necessary, i.e., $y = f(X)$ instead of $f(C, X)$.

\section{Methodological approach}
\label{sec:ml}
This section focuses on the key concepts of the four methodological stages of our study: (i) data preparation, (ii) model implementation, (iii) model's performance evaluation, and (iv) the interpretation of the model outputs\footnote{For more detailed information, see Appendix~\ref{apx:ml} and \citet{provost2013data}.}.

\subsection{Terminology}
We first review the terminology used by social scientists and machine learners. In this section, we establish a link between the naming of concepts in social science and those used in ML (refer to Table \ref{apxtab:terminology} in appendix). In regression, the model is estimated, whereas, in ML, the model is trained~\citep{atheyml19}.
The sample (in-sample) used to estimate the parameters of a model is called the \textit{training set}. ML also uses a \textit{test set} (or a hold-out sample) that is a distinct dataset separated from the training set. Through the learning process, these two types of samples make sure that the model is robust and noise-resistant.

The R-squared is a goodness-of-fit measure that is used in regression models, while accuracy is used in classification models. R-squared is a statistical measure that represents the proportion of variance in the dependent variable predictable from the independent variable, also known as the coefficient of determination. When R-squared value is 1, it shows that the regression prediction model perfectly fits the data. Accuracy is the fraction of prediction that the model measures correctly. It reflects the ability of a model to predict (or classify) classes of unknown vectors since accuracy is generally measured on the dataset that was not used for optimizing the model. Strictly, the metrics are not comparable one to one.

In ML, \textit{features} (variables, columns) refer to the regressors, predictors, or covariates. Each row is called an \textit{example, instance, or observation}.
Our approach is a \textit{supervised learning} approach since both the predictors/features $X_i$ and the output $y_i$ are observed. Another way to categorize a problem is regression and classification. When the output is numerical and continuous, it is called a \textit{regression}. However, when the output is categorical or binary, it is a \textit{classification} problem. Based on the research problem and the dataset we have, we solve a \textit{supervised learning classification} problem.
In the remaining of the paper, we mainly use the ML-based terminologies. Appendix \ref{subsec:methods} describes more information.

\subsection{Data preprocessing}
It is critical to perform data preprocessing before running a model since it removes inconsistencies, missing data, and possible scale/type-related problems. Many ML algorithms only support numerical variables, often for the sake of implementation efficiency. Given a dataset with many categorical variables (e.g., survey questions with yes/no answers), we convert the categorical variables into numerical variables using the \textit{one-hot encoding} method\footnote{A dummy variable that represents categorical data.}. \textit{Discretization} is also typically used to avoid an over-sensitivity of floating numbers, which we used for SPEI drought index values. Section~\ref{sec:data} elaborates on how we build and prepare our dataset.

\subsection{Model implementation: Tree-based approaches}
\label{sec:models}
\label{sec:param:opt}
In this paper, we focus specifically on tree-based methods because these methods because it well suits these methods for classification problems and automatically captures nonlinearity \citep{sendhil17}. As a result, tree-based algorithms are more and more used in applied sciences~\citep{athey2018, atheyml19}.

\begin{figure*}
    \centering
    \resizebox{\textwidth}{!}{
    \includegraphics{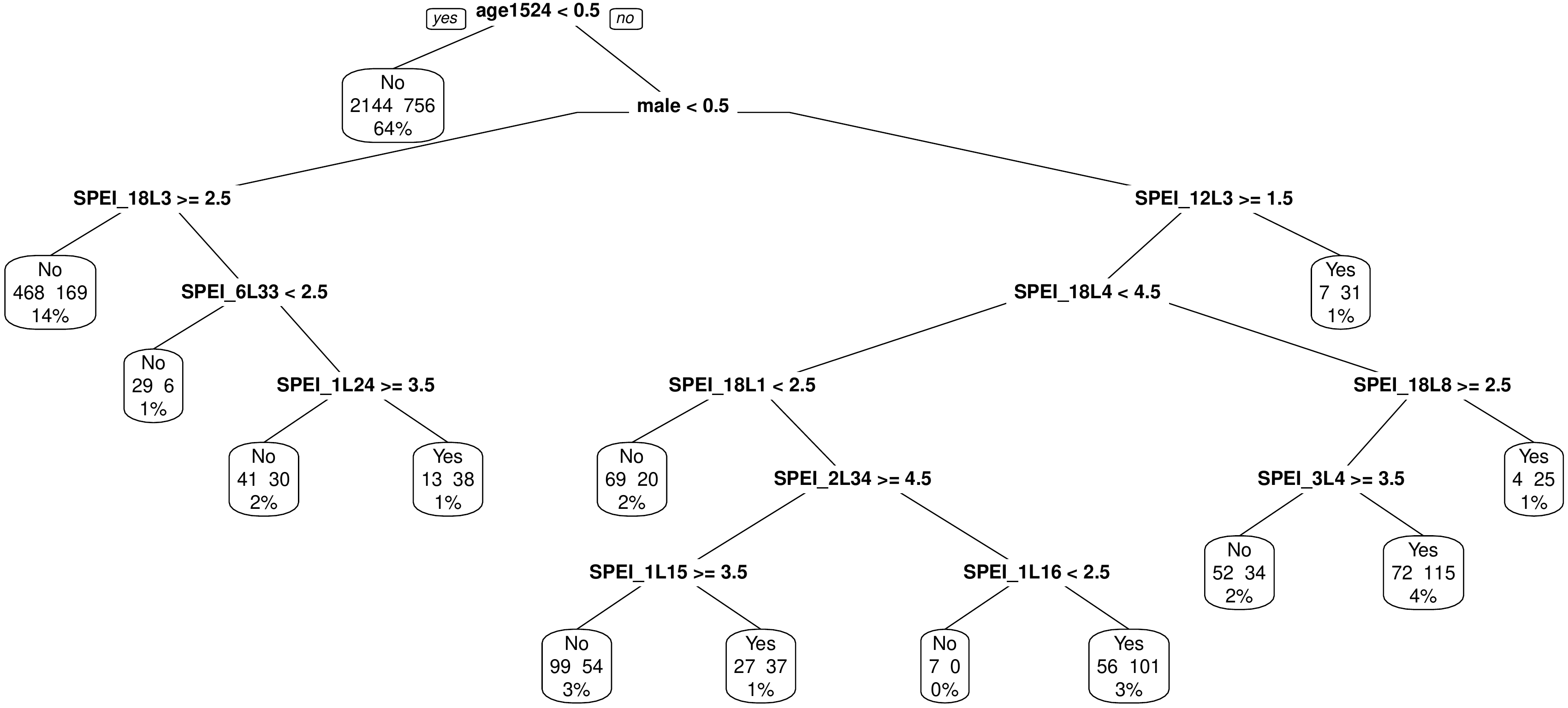}}
    \caption{A decision tree with the features involved in the international moving intention of Burkina Faso.}
    \label{fig:burkbmigdt}
\end{figure*}

The decision tree method consists of approximating the learning function $f$ using decision trees. Figure~\ref{fig:burkbmigdt} is an example of a decision tree from our experiment, which is straightforward and highly interpretable. However, in practice, they can be inaccurate~\citep{DBLP:books/lib/HastieTF09}.
Several other tree-based methods have therefore been proposed. Random Forest (RF)~\citep{Breiman2001} and eXtreme Gradient Boosting (XGB)~\citep{ChenGuestrin2016} are widely used methods. For both RF and XGB, the basic idea is to combine several decision trees to make a prediction. The predictions achieved with multiple trees can then be more accurate, generalizing the data appropriately.

However, obtaining a high-performance and accurate model is not trivial. It involves tuning the model parameters, for which we used the \textit{Bayesian Hyperparameter Optimization} (BHO)~\citep{DBLP:conf/icml/SnoekRSKSSPPA15} to select the proper tuning parameter. Appendix \ref{apx:treeApproach} includes more information on the terminology.

\subsection{Performance evaluation}
\label{sec:metric:auc} 
In supervised learning, models are evaluated by making one-on-one comparisons between the predicted outcome ($\hat{y}$) and the real outcome ($y$). From this comparison, in ML, several metrics are used to evaluate a model. This is a benefit of ML over parameter estimation, which typically relies on the assumptions from the data-generating process to ensure consistency~\citep{sendhil17}.

In this paper, we measure accuracy, precision, recall, and the Area Under the ROC (Receiver Operating Characteristics) curve (AUC)~\citep{swets1988measuring, fawcett2006introduction}. The accuracy is a ratio of correctly predicted observations to the total observations. It is an intuitive performance measure, but only when the dataset is symmetric with a balance between false positive and false negative. It measures the total number of predictions a model gets it correct. But it should be used carefully since, for example, if a model shows high accuracy in an environment with most people not having a disease, the model has a high tendency to falsely predict someone who has a disease. It is why other metrics are considered simultaneously. The precision represents the ratio of correctly predicted positive observations to the total predicted positive observations.  It evaluates how precise a model performs in predicting positive observations and is useful when there are many false positives (e.g., email spam). The recall is the ratio of correctly predicted positive observations to all actual true observations. It evaluates how many actual positives are correctly identified and is useful when there are many false negatives (e.g., fraud detection). However, having a high accuracy (or recall, precision) of a model does not necessarily mean that it is good. It is crucial to use a proper metric for different problems and overview all the metrics. The AUC represents the overall performance of a model regardless of any classification threshold, e.g., $0.5$ to separate positive/Yes ($> 0.5$) and negative/No ($\leq 0.5$) predictions. 
These metrics have values between $0$ and $1$ (the higher, the better performance). Appendix \ref{apx:perfEval} and Figure \ref{fig:confusionmatrixevals} include more information on the terminologies.

\subsection{Output interpretation: Feature importance and Partial Dependence Plots (PDP)}
\label{sec:pdp}
The features $X$ used to estimate $f$ in the equation $y = f(X)$ are rarely equally relevant. 
Typically, only a small subset of features is relevant.  
Hence, after training the model, the \textit{Relative Feature Importance} (RFI) method is used to determine the most relevant ones. It consists of computing the contribution of each feature to the prediction~\citep{breiman1984classification}. 

RFI has become widespread and is thereby used for other ML methods. 
To understand in which direction these important features influence the outcome $y$, \textit{Partial Dependency Plots} illustrates the impact~\citep[Chap. 14]{DBLP:books/lib/HastieTF09}. It is a \textit{marginal average} of $f$ describing the effect of a chosen set of features $S$ on $f$. The most convenient way to compute the partial dependency of a feature $X_i$, that contains $k$ distinct values ($\{x_{i1}, x_{i2},\cdots, x_{ik}\}$), is to compute the prediction when $X_i = x_{ij}$ with $j\in [1,k]$. Appendix \ref{apx:pdp} includes more details.

\section{Data preparation}
\label{sec:data}

\begin{figure}
    \centering
    \resizebox{0.85\columnwidth}{!}{
    \includegraphics{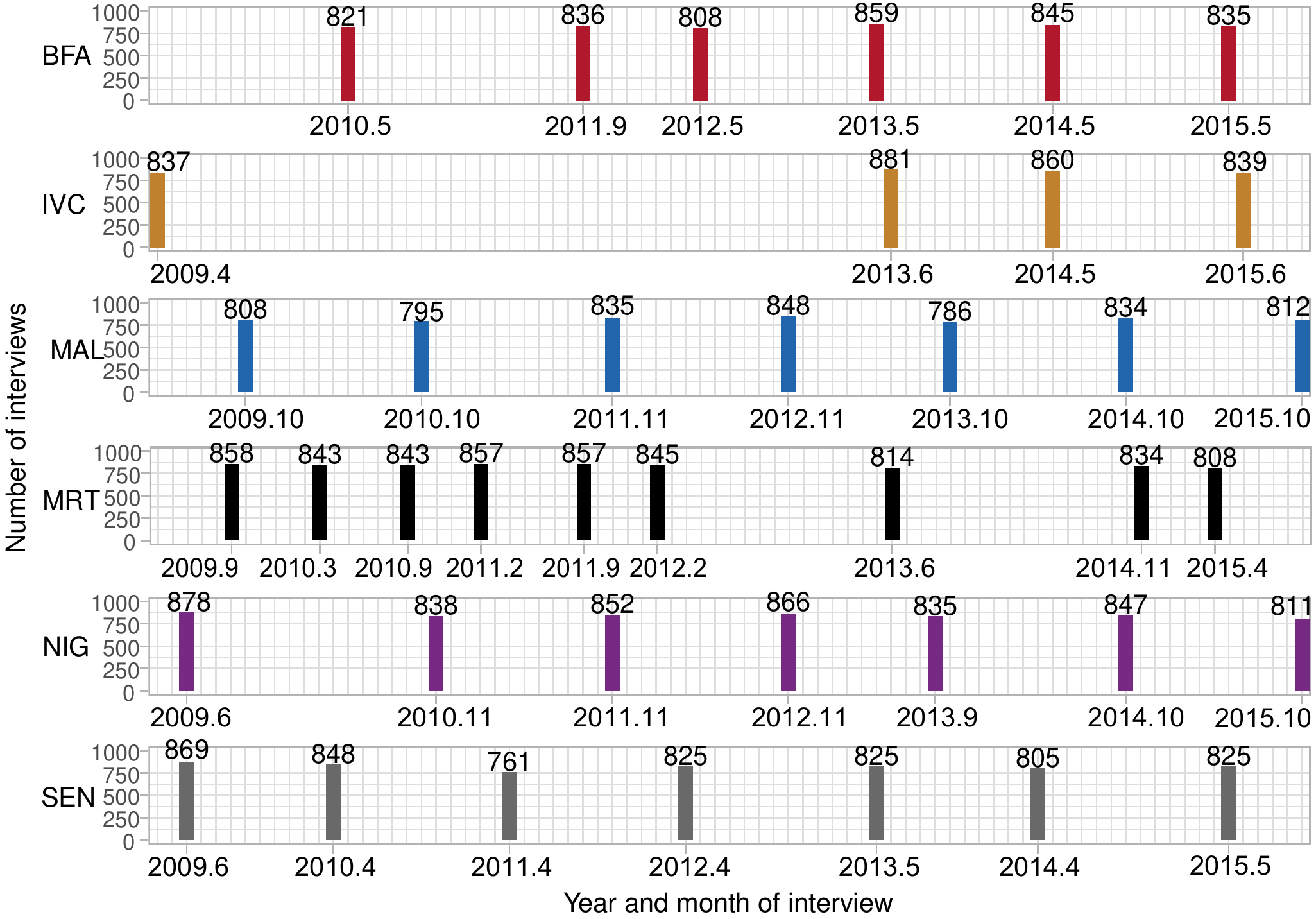}}
    \caption{GWP interview timeline and the number of interviews which are provided in \citet{wthrShock2020}. BFA: Burkina Faso, IVC: Ivory Coast, MAL: Mali, MRT: Mauritania, NIG: Niger, and SEN: Senegal. }
    \label{fig:GWPtimeline}
\end{figure}

In this section, we describe the data sources used in this study and its preprocessing. 
The dataset comprises individual survey data on migration intentions (Section \ref{sec:dataGWP}, Figure \ref{fig:GWPtimeline}) based on Gallup World Poll (GWP)~\citep{gallup2015worldwide} and the weather shocks data based on SPEI (Standardised Precipitation Evapotranspiration Index)~\citep{VicenteSerrano2010} of the six Western African countries between 2009 and 2015 (Section \ref{sec:dataWeather}, Figure \ref{fig:spei55years}). The two datasets are joined by the interviewed months and the region identifiers\footnote{GADM: the Database of Global Administrative Areas} based on a finer geographical identifier which corresponds to the location of an interviewee (i.e., regional administrative units).

\subsection{Gallup World Poll (GWP) data}
\label{sec:dataGWP}

We use the GWP data to study the influences on the likelihood of people who want to move or stay from their country of residence. GWP surveys interview citizens in 160 countries since 2005, covering both urban and rural areas. 
These surveys measure the attitudes and behaviors of a random sample of approximately 1,000 individuals in each survey round. 
Our dataset includes migration responses and other characteristics of interviewees aged from 15 to 49 years from six Western African countries (Burkina Faso, Ivory Coast, Mali, Mauritania, Niger, and Senegal) between 2009 and 2015 (Figures \ref{fig:GWPtimeline} and \ref{fig:datasetsizes}).

\subsubsection{Migration intentions}
\label{sec:dataGWP1}
Two questions have been identified from the GWP survey, and cited by~\citet{wthrShock2020}, which are related to migration intentions.

\begin{itemize}
	\item  \textit{Q1:} In the next 12 months, are you likely or unlikely to move away from the city or area where you live?  (general move)
	\item  \textit{Q2:} Ideally, if you had the opportunity, would you like to move permanently to another country, or would you prefer to continue living in this country? (international move)
\end{itemize}

\textit{Q1}, which we name general move, involves migration including internal and international moving intentions with a decision period of 12 months. Contrary to \textit{Q1}, \textit{Q2} only involves international migration intention excluding a time frame. It should be noted that these questions capture the willingness to emigrate, and one should expect that not all potential migration would realize a move. 

\begin{figure}
    \centering
    \resizebox{0.65\columnwidth}{!}{
    \includegraphics{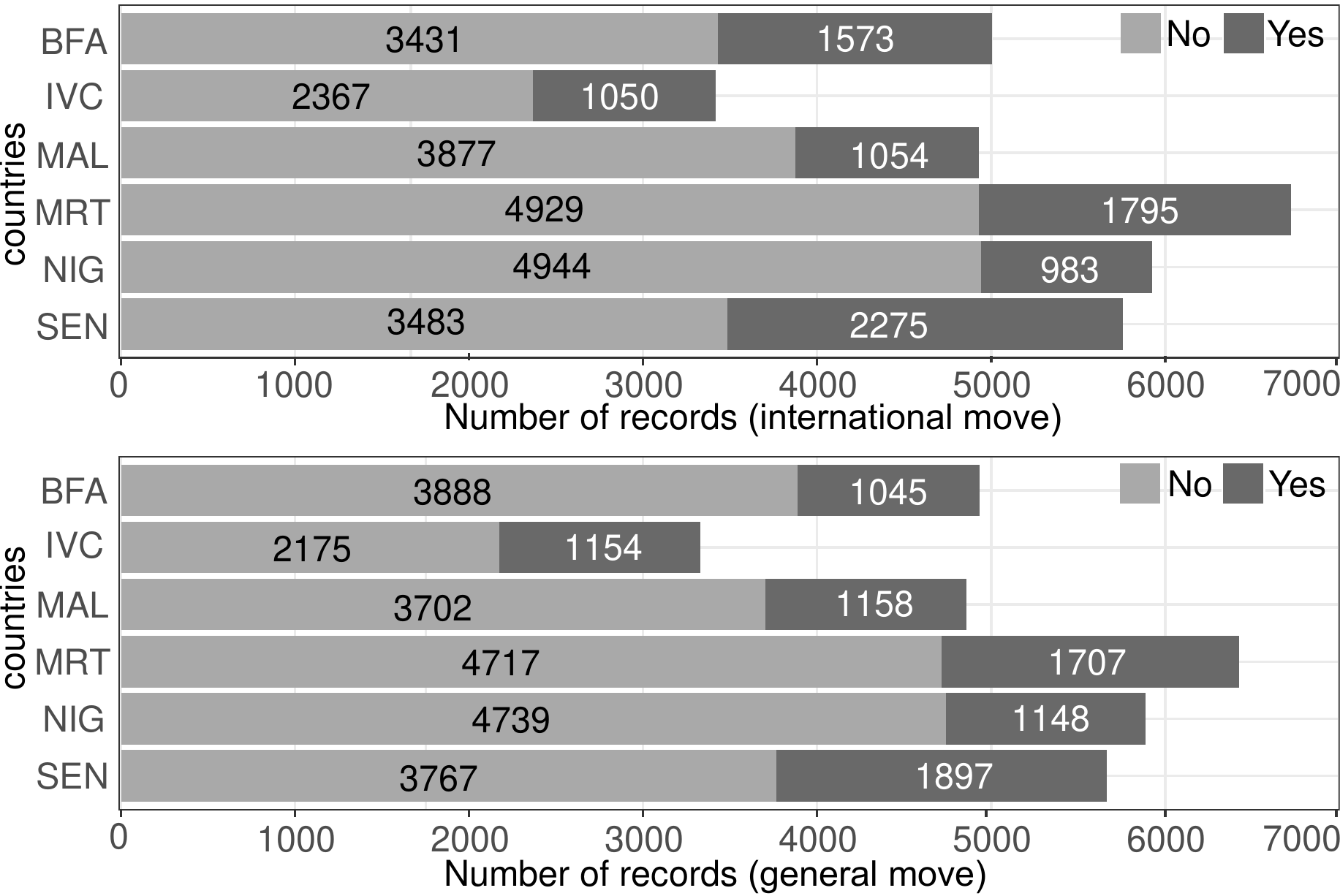}}
    \caption{The number of records in the entire dataset of six countries for international and general move intentions. Refer to Figure \ref{fig:GWPtimeline} for country codes.}
    \label{fig:datasetsizes}
\end{figure}

From these two questions arise the two target variables we aim to explain in our study: general move in \textit{Q1} and international move in \textit{Q2}.  
Figure \ref{fig:datasetsizes} shows the number of records with positive and negative responses for each country towards the two types of moving intentions.

\subsubsection{Individual characteristics}
\label{sec:dataGWP2}

Following the empirical approach of \citet{wthrShock2020}, Table~\ref{tab:datasetvariables} summarizes the control variables used in the ML approach such as country of origin, age, the gender of an individual, and when the interview took place (e.g., month, year). 
Furthermore, `urban' attribute shows whether a person lives in an urban or rural area; `hskill' attribute includes if one is highly educated (i.e., has completed four years of education beyond high school and/or received a 4-year college degree or not); `hhsize' attribute accounts for the number of household members who are older than five years; `mabr' attribute includes whether one has family members or relatives living abroad and who can provide assistance if needed. %
We include `lnhhincpc' attribute, which is the natural logarithm of self-reported household income per capita in dollars. 
This attribute is not included in~\citet{wthrShock2020} because of the side effects it might bring in their identification strategy. %
One of the side effects is that it reduces the sample size, especially since this income question was not asked in all countries. Besides, we cannot overlook the bias introduced because it is a self-reported measurement and that there might be a potential correlation between income and weather shocks~\citep{cattaneo2016migration}. 
This is related to one of the limitations of the traditional empirical approaches discussed in Section~\ref{sec:problem}. 
The reason we include the income variable in our study is that it provides an alternative to the explicit variable selection through the ML approaches. 
We do not make assumptions, but we build the model to select the important variables while being noise-tolerant. %

The preprocessing of the data consists mainly of either binarizing or one-hot encoding certain variables. 
The binary variables `gender', `mabr', `hskill', and `urban' are not concerned in this operation. 
The categorical variables `origin' and `year'\footnote{There are several ways to configure the year variable: (i) use the integer value for each year, (ii) subtract each year by the minimum year to have relatively smaller numbers starting with 0, and (iii) treat integer as a categorical variable and perform one-hot encoding. Here, we use the last approach.} are preprocessed by one-hot encoding. 
The numerical variables `age', `hhsize', and `lnhhincpc' are binarized. 
The age variable is binarized to 15-24 (age1524), 25-34 (age2534), and 35-49 (age35plus)~\citep{wthrShock2020}.

We binarized the variables `hhsize' and `lnhhincpc' into four classes based on a process that tests several subdivisions of the continuous values. It measures the correlation of each class with the dependent variables~\citep{duan2014selecting}\footnote{We used the R package \textit{correlationfunnel} which is fast and offers visualizations to facilitate this work. }. 
The more correlated subdivisions of continuous values are grouped together. For example, variables `hhsize' 3 and 4 (i.e., interviewees who had three and four residents in a house) are grouped as one class, `hhsize 3-4', since they show a high correlation with the dependent variables.

The final GWP dataset, illustrated in Table~\ref{tab:datasetvariables}, consists of six countries variables (origin), seven-year variables (the interview held between 2009 and 2015)\footnote{The interviews are conducted in different months for different countries and the month of interview may be different for each year (Figure \ref{fig:GWPtimeline}).}, four household size variables, four self-reported household income per capita variables, three age variables, gender, living environment (urban or rural), connections abroad (`mabr'), and the individual's education level (`hskill') variables.

\begin{table}
\centering
\resizebox{0.8\columnwidth}{!}{
\begin{tabular}{c|c|ccrr}
\toprule
Type         &       & \shortstack{Feature}   
& one-hot encoding 
& \multicolumn{1}{c}{international}                                                                      & \multicolumn{1}{c}{general}                                                                            \\ \midrule
\multirow{30}{*}{\makecell{GWP\\($X$)}}  & \multirow{30}{*}{\makecell{GWP\\dataset\\only}}    & origin & \begin{tabular}[c]{@{}c@{}}Burkina Faso\\ Ivory Coast\\ Mali\\ Mauritania\\ Niger\\ Senegal\end{tabular} & \begin{tabular}[c]{@{}c@{}}5,004 (16\%)\\ 3,417 (10\%)\\ 4,931 (16\%)\\ 6,724 (21\%)\\ 5,927 (19\%)\\ 5,758 (18\%)\end{tabular}          & \begin{tabular}[c]{@{}c@{}}4,933 (16\%)\\ 3,329 (10\%)\\ 4,860 (16\%)\\ 6,424 (21\%)\\ 5,887 (19\%)\\ 5,664 (18\%)\end{tabular}          \\ \cmidrule{3-6} 
                      &                                      & year      & \begin{tabular}[c]{@{}c@{}}2009\\ 2010\\ 2011\\ 2012\\ 2013\\ 2014\\ 2015\end{tabular}                   & \begin{tabular}[c]{@{}r@{}}4,143 (13\%)\\ 3,975 (12\%)\\ 4,834 (15\%)\\ 4,019 (13\%)\\ 4,835 (15\%)\\ 5,025 (16\%)\\ 4,930 (16\%)\end{tabular} & \begin{tabular}[c]{@{}r@{}}4,037 (13\%)\\ 3,944 (13\%)\\ 4,748 (15\%)\\ 3,951 (13\%)\\ 4,765 (15\%)\\ 4,925 (16\%)\\ 4,727 (15\%)\end{tabular} \\ \cmidrule{3-6} 
                      &                                      & hhsize    & \begin{tabular}[c]{@{}c@{}}inf.-3\\ 3-4\\ 4-6\\ 6-inf.\end{tabular}                                      & \begin{tabular}[c]{@{}r@{}}10,905 (34\%)\\ 5,767  (18\%)\\ 8,933 (28\%)\\ 6,156 (19\%)\end{tabular}                            & \begin{tabular}[c]{@{}r@{}}10,663 (34\%)\\ 5,637 (18\%) \\ 8,745 (28\%) \\ 6,052 (19\%) \end{tabular}                            \\ \cmidrule{3-6} 
                      &                                      & lnhhincpc & \begin{tabular}[c]{@{}c@{}}inf.-5.605\\ 5.605-6.446\\ 6.446-7.231\\ 7.231-inf.\end{tabular}              & \begin{tabular}[c]{@{}r@{}}7,961 (25\%)\\ 7,903 (25\%) \\ 7,935 (25\%) \\ 7,962 (25\%) \end{tabular}                            & \begin{tabular}[c]{@{}r@{}}7,834 (25\%) \\ 7,732 (25\%) \\ 7,780 (25\%) \\ 7,751 (25\%) \end{tabular}                            \\ \cmidrule{3-6}  
                      &                                      & age       & \begin{tabular}[c]{@{}c@{}}age1524\\ age2534\\ age35plus\end{tabular}                                    & \begin{tabular}[c]{@{}r@{}}11,493 (36\%) \\ 10,686 (34\%)\\ 9,582 (30\%) \end{tabular}                                    & \begin{tabular}[c]{@{}r@{}}11,239 (36\%) \\ 10,462 (34\%) \\ 9,396 (30\%) \end{tabular}                                    \\ \cmidrule{3-6} 
                      &                                      & gender    & \begin{tabular}[c]{@{}c@{}}male\\ female\end{tabular}                                                    & \begin{tabular}[c]{@{}r@{}}16,937 (53\%) \\ 14,824 (47\%) \end{tabular}                                            & \begin{tabular}[c]{@{}r@{}}16,593 (53\%) \\ 14,504 (47\%) \end{tabular}                                            \\ \cmidrule{3-6}  
                      &                                      & urban     & \begin{tabular}[c]{@{}c@{}}urban\\ rural\end{tabular}                                                    & \begin{tabular}[c]{@{}r@{}}7,491 (24\%) \\ 24,270 (76\%) \end{tabular}                                             & \begin{tabular}[c]{@{}r@{}}7,295 (23\%) \\ 23,802 (77\%) \end{tabular}                                             \\ \cmidrule{3-6}  
                      &                                      & mabr      & \begin{tabular}[c]{@{}c@{}}yes\\ no\end{tabular}                                                         & \begin{tabular}[c]{@{}r@{}}14,879 (47\%) \\ 16,882 (53\%) \end{tabular}                                            & \begin{tabular}[c]{@{}r@{}}14,573 (47\%) \\ 16,524 (53\%) \end{tabular}                                            \\ \cmidrule{3-6} 
                      &                                      & hskill    & \begin{tabular}[c]{@{}c@{}}yes\\ no\end{tabular}                                                         & \begin{tabular}[c]{@{}r@{}}886 (3\%) \\ 30,875 (97\%) \end{tabular}                                              & \begin{tabular}[c]{@{}r@{}}862 (3\%) \\ 30,235 (97\%) \end{tabular}                                              \\ \midrule
\multirow{2}{*}{\makecell{SPEI \\($C$)}} & 
\multirow{2}{*}{\makecell{SPEI\\dataset}} &  SPEI timescales    & 1,2,3,6,12,18,24    &   &        \\ \cmidrule{3-6} 
    &   &  lags  &  lag0 - 48  &      & \\ \midrule
ALL &     &   GWP + SPEI        &      &   & \\ \bottomrule
\end{tabular}%
}
\caption{The number of samples of the binarized and discretized dataset with one-hot encoding for the international and the general move.}
\label{tab:datasetvariables}
\end{table}

\subsection{Weather shocks data} %
\label{sec:dataWeather}

The results from~\citet{wthrShock2020} show that an identified period of shocks, the intensity of the shocks, and the treatment of the (local) crop-growing or crop-planting seasons have impacts on the migration intentions (general and international). This section describes the weather shock information we used.

\subsubsection{SPEI}
\label{sec:dataWeatherSPEI}
The economic activity of the region we focus on highly depends on the agricultural sector. In the absence of irrigation infrastructure, weather, and in particular, water availability directly influences agricultural production. In such a context, livelihoods are indirectly affected by the climatic condition. One strategy to deal with chronic weather variability, especially when other economic opportunities are limited, is to move.

One of the statistical challenges in the literature studying the impact of climate variability on various economic outcomes, including migration, is how to measure it so it is comparable over time, space, and several relevant climatic factors. At earlier stages, literature has focused on precipitation. However, the impact of climate on agricultural yields depends on factors such as temperature and the ability of the soil to keep water. Moreover, the emerging global warming issue emphasizes the importance of capturing the impact of temperature. These are assembled by the potential evapotranspiration (PET), which in turn depends on temperature, latitude, sunshine exposure, and wind speed.

\begin{figure}%
   \centering
   \resizebox{0.9\columnwidth}{!}{
            \includegraphics[scale=1.0]{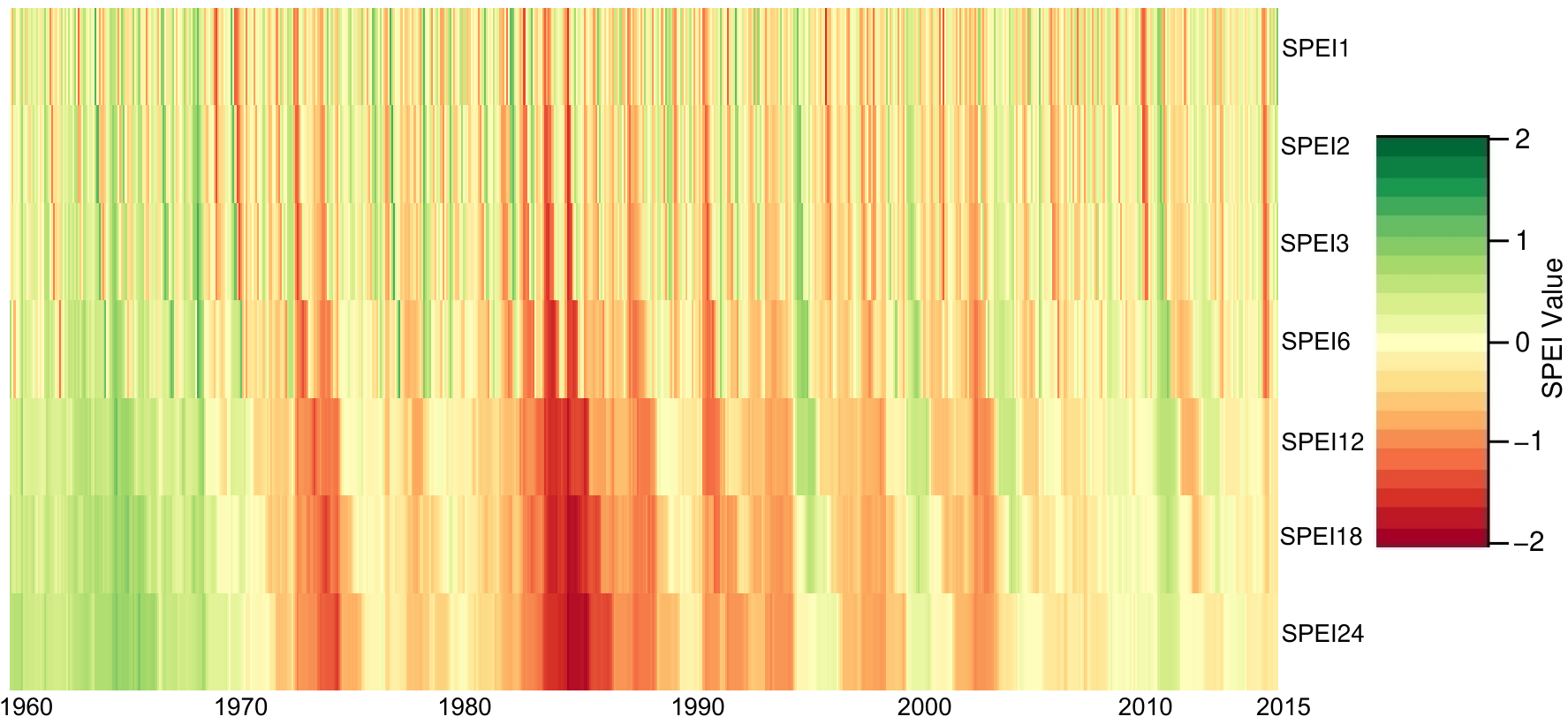}
   }
   \caption{SPEI timescales for 55 years in six western African countries.} 
   \label{fig:spei55years}   
\end{figure}

SPEI is calculated by fitting the time series of differences of precipitation and PET (i.e., climatic water balance) to a probability distribution. This process enables the differences as standard normal scores with zero mean and unit variance. It is standardized using a Log-logistic distribution function which is found to be the most suitable distribution for SPEI \citep{vicente2010new}. These standardized units are comparable in space, time, and different SPEI timescales. The index value below zero is characterized as a drought. These calculated monthly SPEI values are collected at different time scales for each subregion in the six western African countries. Figure \ref{fig:spei55years} illustrates these SPEI values averaged over the six countries between 1960 to 2015 to identify the dry and wet conditions in this area. It explicitly shows moderate conditions before 1970 in green, while increasing drought shocks trend with SPEI at 24 timescales in the early 1970s and between 1980s and 1990 in red.

 SPEI outperforms other indices in predicting agricultural yields \citep{vicente2010new}, especially as an index incorporating the effect of temperature, which can assess the emerging global warming effects. In fact, SPI (Standard Precipitation Index) and SPEI are similar in the way of calculating the index, but SPEI overcomes the limitation of SPI by comparing the water and the evapotranspiration in the atmosphere. Unfortunately, SPI considers only the precipitation. Both SPI and SPEI are multi-scaled indices\footnote{SPEI at 3 months timescale for May 2015 is a function of the sum of the climatic water balance of March, April, and May 2015.} that can identify the multi-temporal nature of droughts. Another advantage of the SPEI is that it considers the onset, length, and intensity of a climatic event, rather than only the absolute value of precipitation and temperature (Figure \ref{fig:monthlyInterview}). Moreover, it is comparable over time and space thanks to its fixed mean and standard deviation\footnote{By construction, SPEI has a zero mean and a standard deviation of unity.}.

\subsubsection{SPEI and Lags}
\label{sec:dataWeatherSeverityLags}

The SPI, and therefore SPEI, was initially designed to quantify the precipitation deficit for multiple timescales. Such timescales reflect the incidence of drought on the water source availability. The climatology community has defined three main types of drought: (i) meteorological drought, (ii) hydrological drought, and (iii) agricultural drought which differ in intensity, duration, and spatial coverage (\citet{eslamian2014handbook}). The meteorological drought captures the extent to which soil moisture conditions react to precipitation anomalies in the short run, whereas surface and groundwater reservoirs are subject to the longer-term precipitation anomalies as captured by hydrological droughts. Agricultural drought occurs when crop production is affected by precipitation anomalies. To some extent, the meteorological drought is the mildest scenario, whereas the hydrological drought is the most severe scenario of drought occurrence, with the agricultural drought being in between. In this sense,  a 1 or 2-month SPEI measure can show a presence and level of meteorological drought; from 1 month to 6-month SPEI for agricultural drought, and from 6 months to 24-month SPEI or more for hydrological droughts (\citet{wilhite2000drought}). 

Our key challenges are to understand which measure of climate conditions matters for migration and which time-spans (or timescales) need to be considered. 
We built the weather shocks dataset with seven SPEI timescales (i.e., 1, 2, 3, 6, 12, 18, 24). We collect the SPEI values based on each subregion and an interview month. 
We gather the lags of each timescale of SPEI from the past four years from the interview month, which makes a total of 49 lags (lag0--lag48)\footnote{To get a SPEI at 12 months timescale with lag 6 for an individual interviewed in May 2015, the SPEI value is the SPEI12 value 6 months ago in November 2014.}. To understand how SPEI drought index affects the migration intention (positive or negative), the continuous values of SPEI are discretized into \textit{seven equal bins}. For a feature $X_i$, the binning step is therefore
\begin{equation}
    \text{bin step}(X_i) = \frac{max(X_i) - min(X_i)}{7}
\end{equation}
where $i$ is the SPEI timescale. This discretization allows limiting the sensitivity of our models to the high variability of SPEI values. Figure \ref{apxfig:speiDiscretization} in the appendix provides an example of discretization.

\section{Results}
\label{sec:results}
In this section, we report the results by applying machine learning algorithms to the constructed dataset combining the migration intention data (GWP) and the weather shock data (SPEI) (see Section \ref{sec:data}).

\subsection{Protocols}
\label{sec:protocol}
We ran all experiments on a computing environment with an Intel Core i5 64-bit processor (2.7GHz) and 16GB of RAM running MacOS. 
We ran tests on the entire dataset (ALL) and then on the feature groups described in Table~\ref{tab:datasetvariables}. To reduce the risk of overfitting, we use the 10-fold cross-validation workflow.
For each of these datasets and each dependent variable (general or international migration intention), we conducted experiments separately using the R implementation of the tree-based ML algorithms: DT: Decision trees (\citet{Quinlan1986}), RF: Random Forest (\citet{Breiman2001}), and XGB: eXtreme Gradient Boosting (\citet{ChenGuestrin2016}). Table~\ref{tab:param} shows the optimal parameters (Section~\ref{sec:param:opt}) used for these algorithms. Our findings follow in the order of questions asked in Section \ref{sec:questions}.

\begin{table}[!htb]
    \centering
    \begin{tabular}{ccc}
    \toprule
         DT &  RF & XGB \\ 
         \midrule
         cost complexity = $10^{-5}$ & mtry = 5 & mtry = 3 \\
         tree depth = 30 & number of trees = 1080 &  number of trees = 761 \\
         number of nodes = 20 & & \\
         \bottomrule
    \end{tabular}
    \caption{Optimal parameters for each algorithm. \textit{mtry}: the number of possible splits at each node.}
    \label{tab:param}
\end{table}

\subsection*{\textit{Q1}. Performance comparison of tree-based ML algorithms }

\begin{figure}
   \centering   
   \includegraphics[scale=0.7]{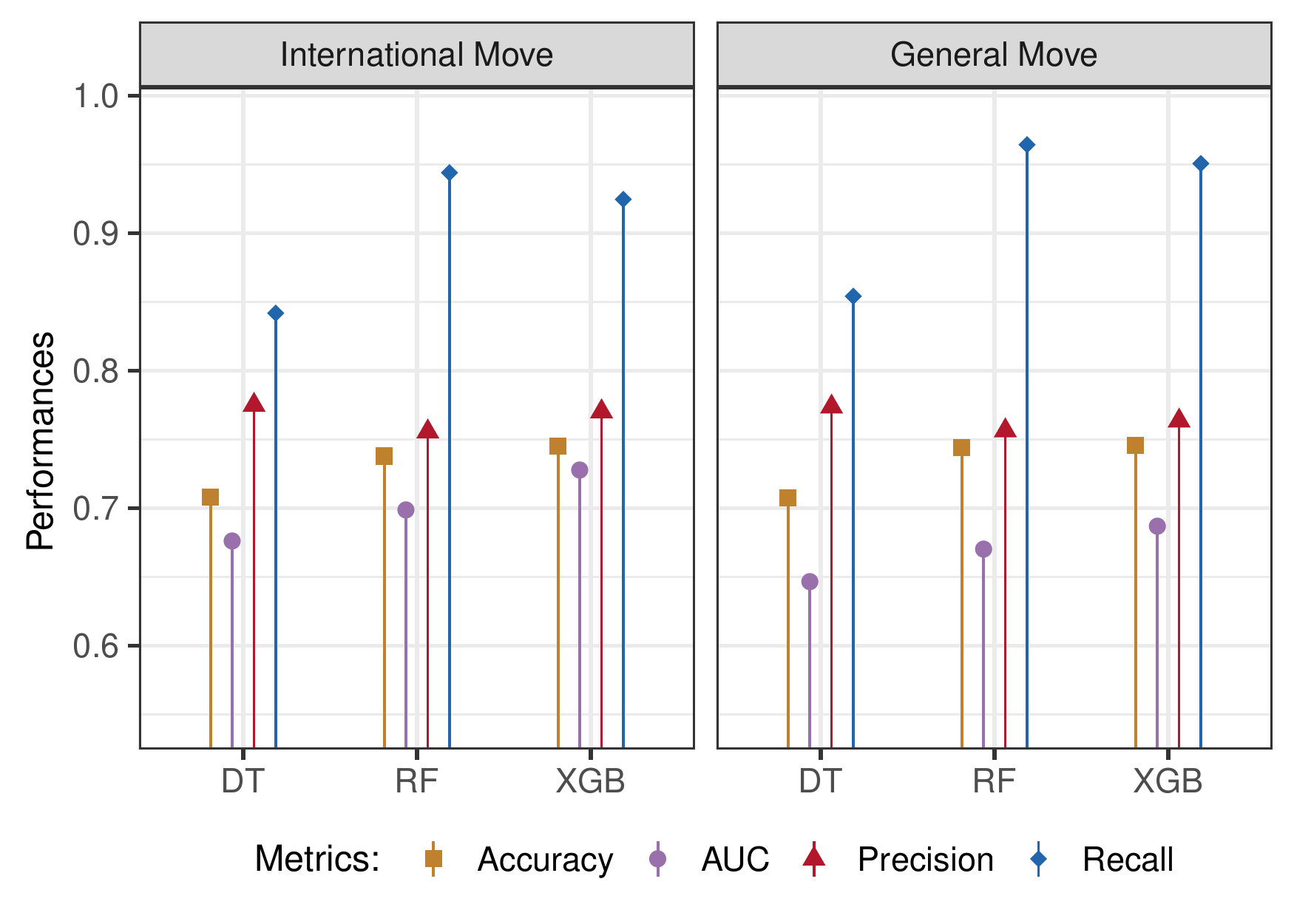}
   \caption{Performance of tree-based algorithms of the test set in the entire dataset (i.e., GWP and Weather data) for all six countries.} 
   \label{fig:all:algo}   
\end{figure}

We compare the results of tree-based methods over the test set with DT, RF, and XGB in Figure~\ref{fig:all:algo} based on AUC (refer to Section~\ref{sec:metric:auc}). The AUC measure is used to compare the models since it characterizes the overall performance of the classifiers.
We find that XGB outperforms other algorithms for both dependent variables. Thus, the results described afterward are all from the XGB algorithm.

\begin{table}%
\centering
\begin{tabular}{clrrrrl}
\toprule
Metric & move &  mean.ALL  &  mean.GWP  &  t.value  &  p.value  &  comparison \\
\midrule
\multirow{2}{*}{Precision} & International & 0.77 & 0.76(0.7567) & 4.00 & 0.00 & $\text{ALL} > \text{GWP}$ \\ 
& General & 0.76 & 0.75(0.7462) & 4.76 & 0.00 &  $\text{ALL} > \text{GWP}$ \\ 
\midrule
\multirow{2}{*}{AUC} &   International & 0.73 & 0.71 & 5.51 & 0.00 & $\text{ALL} > \text{GWP}$ \\
 & General & 0.69 & 0.66 & 7.70 & 0.00 & $\text{ALL} > \text{GWP}$ \\
\midrule
\multirow{2}{*}{Accuracy} & International & 0.75 & 0.74(0.7416) & 1.95 & 0.08 & $\text{ALL}$ \textasciitilde{} $\text{GWP}$ \\ 
  & General & 0.75 & 0.74(0.7384) & 1.93 & 0.08 & $\text{ALL}$ \textasciitilde{} $\text{GWP}$ \\ 
\midrule
\multirow{2}{*}{Recall} & International & 0.92 & 0.95(0.9487) & -10.29 & 0.00 & $\text{ALL} < \text{GWP}$ \\ 
  & General & 0.95 & 0.98(0.9793) & -14.25 & 0.00 & $\text{ALL} < \text{GWP}$ \\ 

\bottomrule
\end{tabular}
\caption{The t-test of GWP compared to ALL (GWP + weather) using the XGB algorithm on test sets.}
\label{tab:ttestallgwp}
\end{table}

\subsection*{\textit{Q2}. Influence of weather towards moving intentions}

To answer the impact of weather towards migration intentions, we conducted separate experiments on the individual survey dataset (i.e., GWP) and the entire dataset with individual survey and weather dataset of the six countries (i.e., ALL = GWP + Weather) using the XGB algorithm. The \textit{t-test} comparisons of precision and AUC measures on the test set reveal that the classifier on the entire dataset (i.e., ALL includes the weather information) significantly outperforms the GWP's classifier for both international and general move dependent variables (Table~\ref{tab:ttestallgwp}). Thus, including weather data improves the prediction power for precision and AUC.  

The t-test comparison in terms of accuracy shows that the models are not different between the individual survey (i.e., GWP) and the entire datasets (i.e., ALL = GWP + Weather). For recall, the GWP's model outperforms the ALL's model, although in general, ALL's model outperforms GWP's model (which is noisier). This would be due to the presence of several countries that do not have the same weather conditions. We investigate this in the following section.

\subsection*{\textit{Q3}. A general model or a country-specific model}

To decide whether we can achieve a general prediction model for all six countries, we investigate further the prediction performances of six countries (general model) and those of each country (country-specific model) on international and general migration intentions. 

\begin{figure}
  \centering
  \resizebox{0.8\columnwidth}{!}{
      \includegraphics{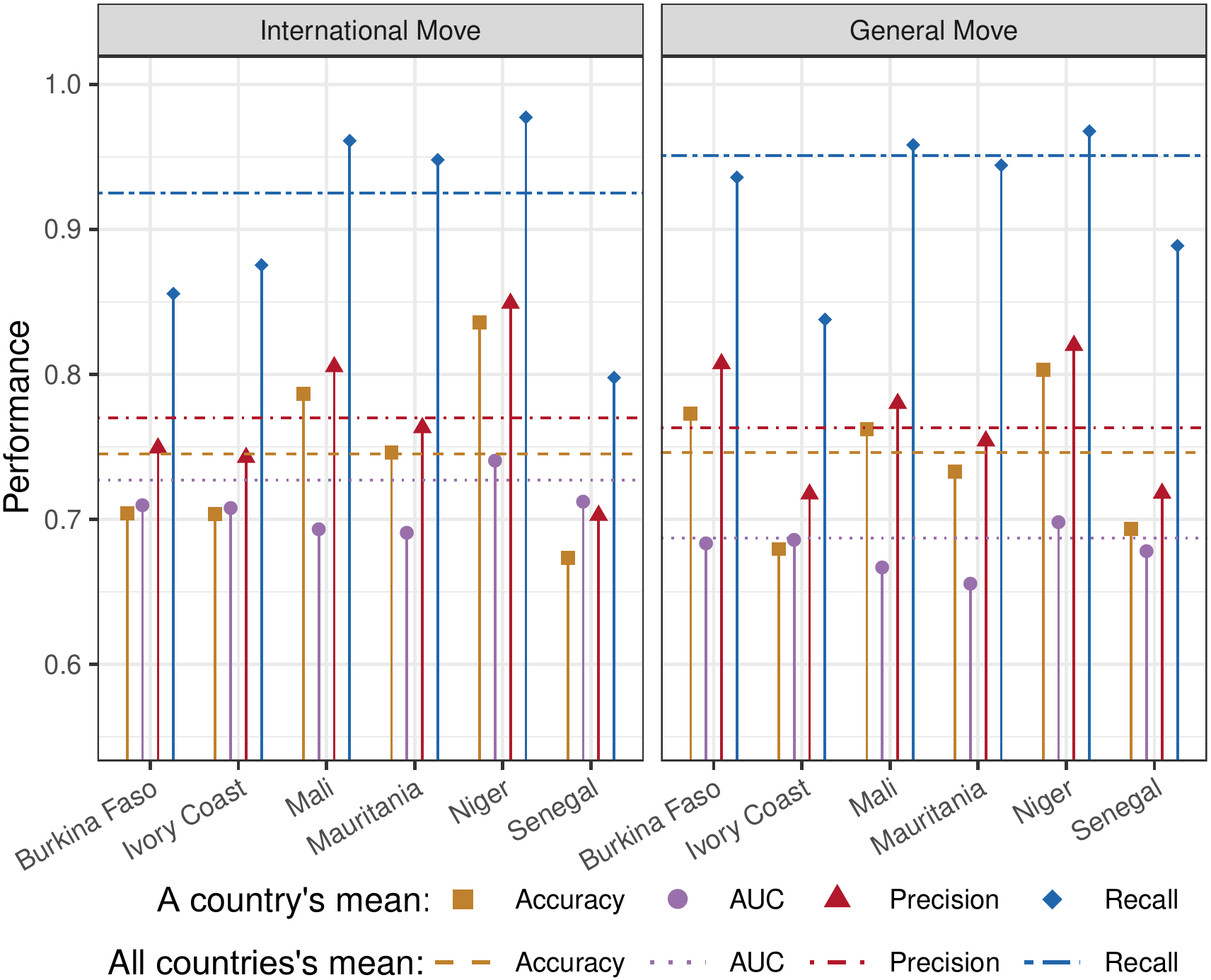}
  }
  \caption{Comparison between the average performance of six-countries' model (dotted lines) and performances of each country model (shape) on the test set. \label{fig:ttestallcountries123}} 
\end{figure}

Figure~\ref{fig:ttestallcountries123} shows compares multiple measures using XGB of each country compared to the mean value with all-countries for the GWP $+$ weather dataset (ALL). The horizontal dashed line for each metric (accuracy, AUC, precision, and recall) represents the average performance of the general model trained with an all-countries dataset. Each vertical line with shape represents the country-specific performance for each metric. For example, considering precision, country-specific models for Mali and Niger outperform the general model using the entire dataset for the international move prediction and the same with Burkina Faso, Mali, and Niger for the general move prediction.

Hence, generalizing moving intentions with one model can mislead since some countries do not fit into a general model. Moreover, the performance can be different for different targets in a country. For example, the precision performance of the general move intention of Burkina Faso's one-country model is higher than the model for all countries; however, it shows lower performance with the international move intention. This guides us it is critical to investigate the model for each country when analyzing the relationship between climate and migration.

\subsection*{\textit{Q4}. Important features}

\begin{figure}%
   \centering
       \includegraphics[width=0.75\textwidth]{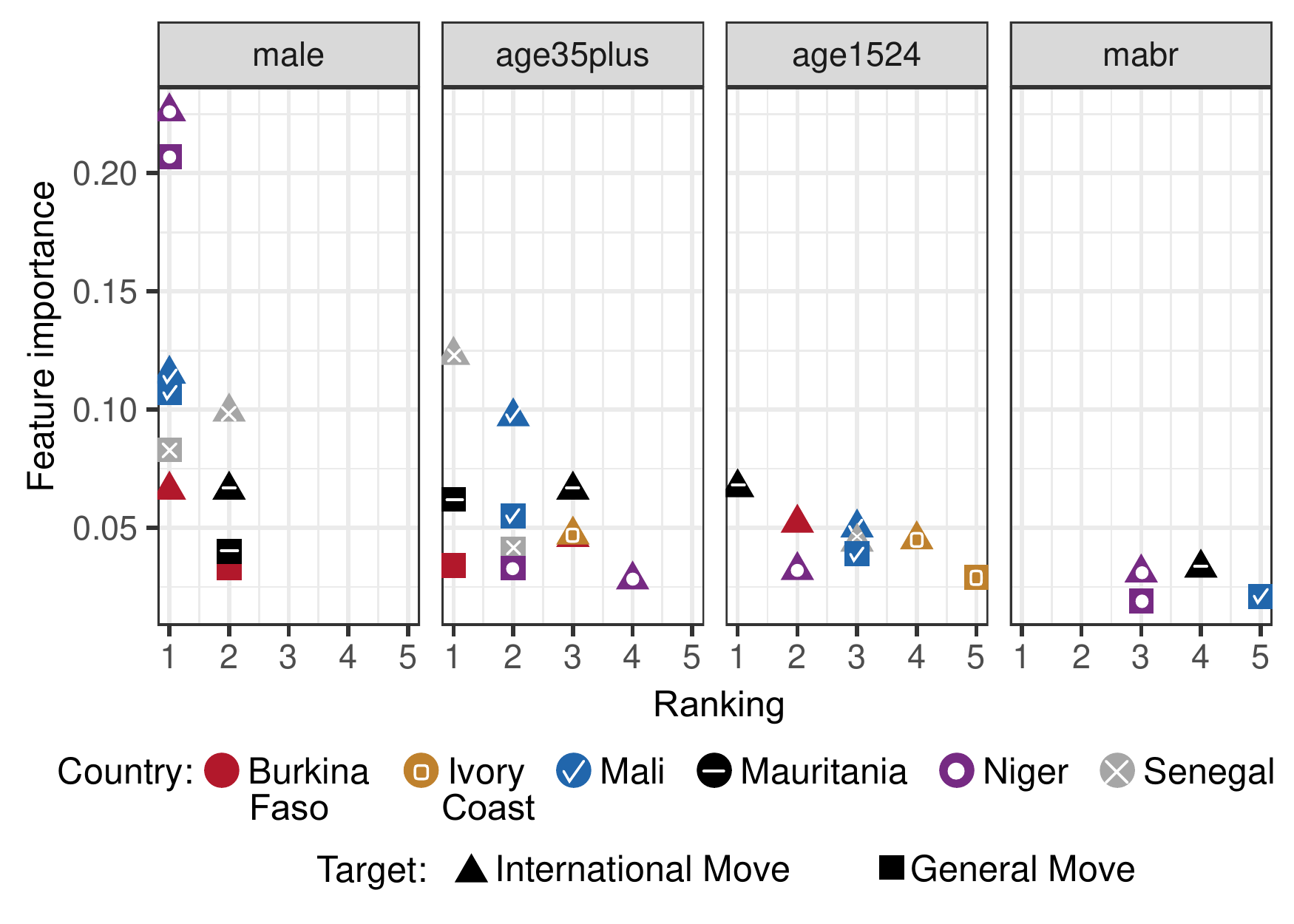}
   \caption{GWP features with higher feature importance on each target.} 
   \label{fig:higherFeatureImp}   
\end{figure}  

For five out of six countries, we find that gender and age show a stronger influence on both the international and general moving intentions\footnote{We find that the results are similar to permutation feature importance. Refer to Figure \ref{apxfig:pfi} in the appendix.}. In Figure \ref{fig:higherFeatureImp}, gender feature (`male') is considered as the first important feature for three countries, respectively on the international and general move intentions. Overall, men have higher intentions to move than women, while people between 35 and 49 are likely to stay at their current residence. While we cannot infer the exact reasons, we assume that males have higher spatial mobility to search for a job and a better economic status. Also, elder adults over 35 are lesser inclined to move, as stated in previous studies (\citet{Djamba2003}).

Besides, the younger group, aged between 15 and 24, are more likely to engage in international migration. Moreover, having a one-distance connection abroad (`mabr') shows a positive impact on migration decisions. SPEI features are the most important features in Ivory Coast, followed by age characteristics. However, one cannot generalize these findings due to the limited number of realized surveys (Refer to Table \ref{tab:datasetvariables}). Interestingly, Niger and Mali are countries that show higher accuracy, precision, and recall measures at Q3, which both have feature importance values higher than $0.1$ on both international and general move intentions (Figure \ref{fig:higherFeatureImp}).
The next section shows the influence of weather features with slightly lower feature importance values. The countries for which weather features show a higher value than the average importance values for each country are Ivory Coast and Mauritania. This finding is relevant for both types of movement, international and general move. With Senegal, weather features show a higher effect than the average importance for the international move intention.

\begin{figure}%
   \centering
       \includegraphics[scale=0.7]{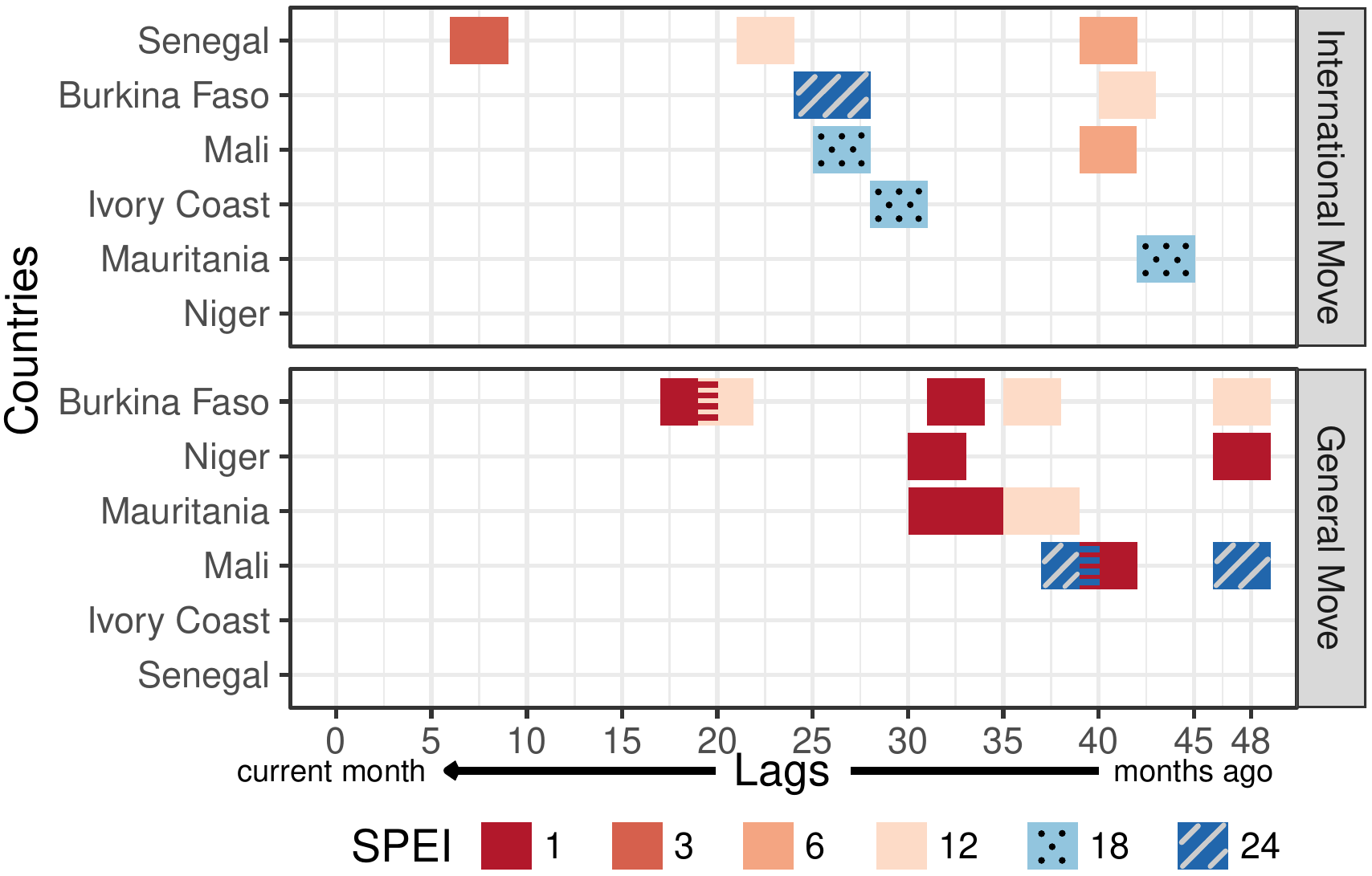}
   \caption{Important SPEI timescales and periods of lags. The order of countries is based on the emergence of important lag periods.} 
   \label{fig:speiLagSummary}   
\end{figure}

\subsection*{\textit{Q5}. Influence of SPEI and monthly lags towards moving intentions}
\label{sec:question5}

Figure \ref{fig:speiLagSummary} shows the most prominent SPEI timescales and lag periods, for at least three consecutive months, based on the feature importance values greater than the average. Each row (y-axis) represents a country and each column (x-axis) shows the lags. The reddish periods represent SPEIs comprising shorter timescales (i.e., 1, 2, 3, 6, 12) while bluish representation for longer ones (i.e., 18, 24)\footnote{Longer timescales ($\geq$18 months) referred to \url{https://climatedataguide.ucar.edu/climate-data/standardized-precipitation-evapotranspiration-index-spei}.}. We find that more reddish plots are more visible in the general move intention and more bluish ones for the international moving intention. This means the international move has more influence from the longer timescales of SPEI as it may involve a more extended period to make permanent decisions\footnote{The international move's question (\textit{Q2}) actually asks people if they want to move permanently to another country.}, while general move which includes internal move, is more influenced by shorter timescales of SPEIs. Especially, Burkina Faso and Mauritania show shorter SPEIs affecting the general moving intentions and Mauritania's international migration intention is affected by longer SPEIs. With lags, besides some periods in Senegal and Burkina Faso, we find that lags over 24 months are more likely to affect migration intentions. Interpreting such a result would be that potential migrants move internally because meteorological/agricultural droughts occur, whereas international migration is a response to more severe, hydrological droughts. This shows that the severity of climate conditions determines the extent to which migration plans would be drastic. As severe weather conditions are highly correlated over time and space, migration is from a longer-distance type (e.g., international migration) to be protected from these conditions. 

It is challenging to draw global patterns since the results are country- and migration-type-specific, thus, further research must consider the heterogeneities of such country and migration type. As mentioned in the previous section regarding the important features, demographic characteristics are the essential drivers for both international and general migration intentions (again considering heterogeneities among countries and migration type). SPEI drought index is important to a lower extent, but seems to explain a part of the migration probability; however, the explanations are different across countries and migration decisions. In Appendix \ref{apx:additional}, Figure \ref{apxfig:speiLag} shows the feature importance in size for each SPEI and lag combination for each country. It includes more details of the distribution of each SPEI index and each lag and the overall results of SPEI timescales and lags combinations.

\begin{figure}
   \centering
       \includegraphics[width=0.75\textwidth]{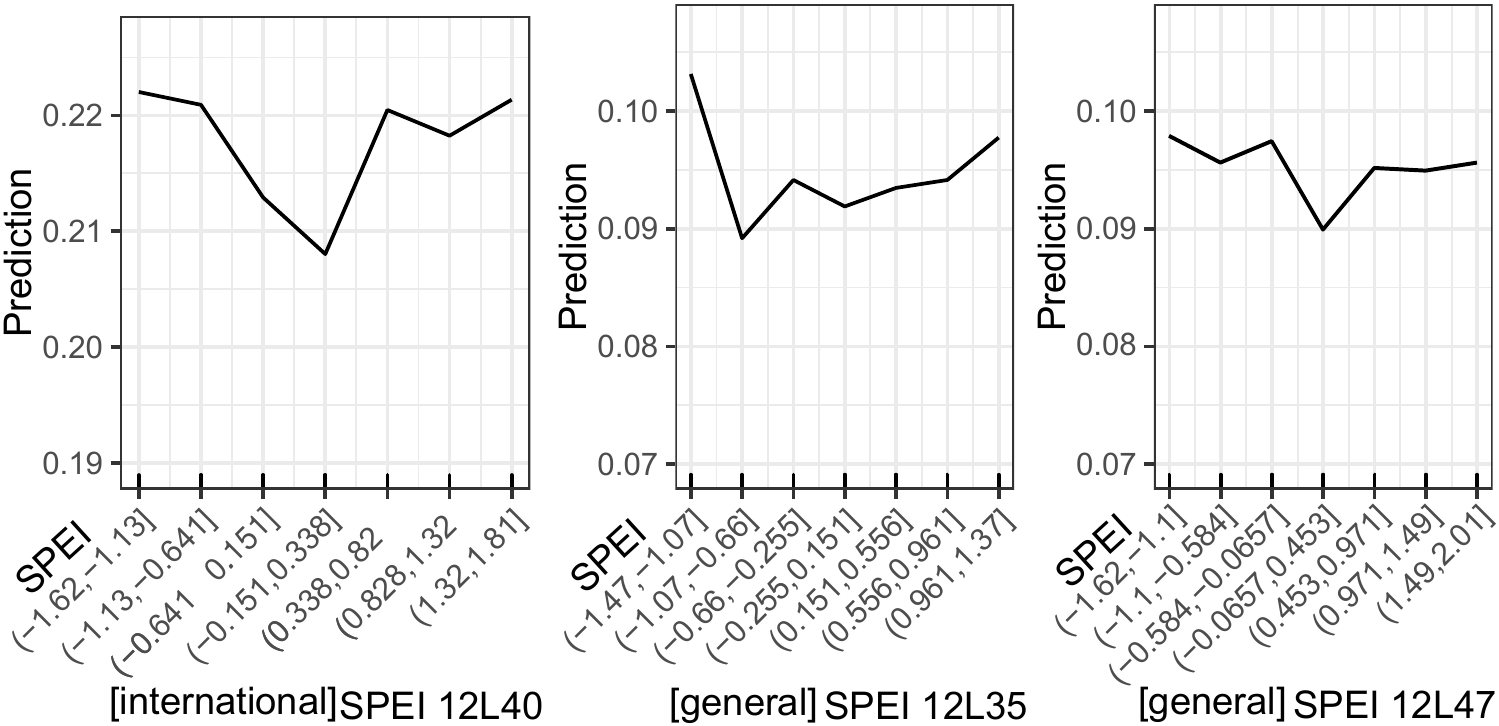}
       \label{brk_bmig_move}   
   \caption{Partial dependence plot (PDP) of selected SPEI timescale and lag combinations with V-shape in Burkina Faso. Severe drought and flooding are indicators of international and general migration intentions.}
   \label{fig:vshape}   
\end{figure}

\begin{figure}
\centering
 \newlength\yearposx
\begin{tikzpicture}[scale=0.615] %
 \foreach \x in {0,...,48}{
        \pgfmathsetlength\yearposx{(\x-0)*0.25cm};
        \coordinate (y\x)   at (\yearposx,0);
        \coordinate (y\x t) at (\yearposx,+3pt);
        \coordinate (y\x b) at (\yearposx,-3pt);
    }
    \draw [-] [draw=red, ultra thick] (y0) -- (y3);
		\draw [-] [draw=green, ultra thick] (y3) -- (y4);
		\draw [-] [draw=brown, ultra thick](y4) -- (y8);
		\draw [-] (y8) -- (y11);
		\draw [-] [draw=red, ultra thick] (y11) -- (y15);
		\draw [-] [draw=green, ultra thick] (y15) -- (y16);
		\draw [-] [draw=brown, ultra thick](y16) -- (y20);
		\draw [-] (y20) -- (y23);
		\draw [-] [draw=red, ultra thick] (y23) -- (y27);
		\draw [-] [draw=green, ultra thick] (y27) -- (y28);
		\draw [-] [draw=brown, ultra thick](y28) -- (y32);
		\draw [-] (y32) -- (y35);
		\draw [-][draw=red, ultra thick] (y35) -- (y39);
		\draw [-] [draw=green, ultra thick] (y39) -- (y40);
		\draw [-] [draw=brown, ultra thick](y40) -- (y44);
		\draw [-] (y44) -- (y47);
		\draw [-][draw=red, ultra thick] (y47) -- (y48);

    \foreach \x in {0,...,48}
        \draw (y\x t) -- (y\x b);
				
		\foreach \x in {0,12,24,36,48}
        \node at (y\x) [below=3pt] {\x};
				
		\foreach \x in {35, 47}
				\node[align=center] at (y\x) [above=1pt] {\textcolor{red}{\x}};
				
		\foreach \x in {40}
				\node[align=center] at (y\x) [above=1pt] {\textcolor{brown}{\x}};

    \foreach \x in {0,12,24,36,48}
        \node at (y\x) [below=3pt] {\x};
				
		\node[align=center, font=\fontsize{6pt}{6pt}\selectfont] at (-.2,.5){\textcolor{red}{planting}};
		\node[align=center, font=\fontsize{6pt}{6pt}\selectfont] at 
		(1.2,.5){\textcolor{dartmouthgreen}{growing}};
		\node[align=center, font=\fontsize{6pt}{6pt}\selectfont] at (2.7,.5){\textcolor{brown}{harvesting}};
		
		\node[align=center, font=\fontsize{8pt}{8pt}\selectfont] at (0.5,-1.4){\textcolor{blue}{interview in May}};

	   	\node[align=center, font=\fontsize{8pt}{8pt}\selectfont] at (0.3,-1){\textcolor{black}{current month}};
	    \node[align=right, font=\fontsize{8pt}{8pt}\selectfont] at (11,-1){\textcolor{black}{months ago}};
	   	
		\draw[black,thick,<-] (2,-1) -- (9.5,-1);
	\end{tikzpicture}
	
\caption{Cropping seasons of sorghum, maize, and millet in Burkina Faso, including the planting growing and harvesting periods over the four years preceding the GWP interview mainly held in May (Figure \ref{fig:GWPtimeline}). Months 35, 40, and 47 show the significant lags presented in Figure \ref{fig:vshape}.}	
\label{fig:monthlyInterview}
\end{figure}
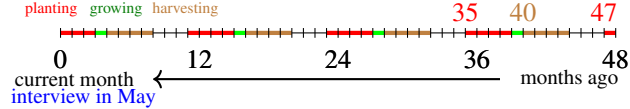

To understand the mechanisms behind the results included in Figure \ref{fig:speiLagSummary}, we further look at: (i) the way climate shock events are captured by the different SPEI indicators influencing migration and (ii) why these lags appear to be important. It is important to note that the results differ among the different countries for the different international and general migration intentions. For example, our findings are explainable for Burkina Faso where the impact of important SPEI is V-shaped (Figure \ref{fig:vshape}), compared to other countries. This indicates that highly negative (e.g., severe drought) and positive values (e.g., severe flooding) of SPEI increase the probability of moving both internationally and globally, whereas values closer to zero reduce it. We observe this V-shaped impact mainly for the months that fall in the cropping season (Figure \ref{fig:monthlyInterview})\footnote{The economic activity in the countries we consider in this article highly depends on the agricultural sector. Knowing that irrigation infrastructure is lacking and that those sectors are mainly rain-fed, weather conditions are important contributors to agricultural production and income generation.}. The cropping season in Burkina Faso that concerns its main crops, sorghum, maize, and millet starts in April and ends with the harvesting period in December. The lack or surplus of rainfall in the Aprils of three and four years, corresponding to lags 35 and 47, before the interview increases the intentions of moving generally and internationally. It is natural to expect that individuals do not immediately react to current events, but rather consider past weather events that occur in periods significant for economic activity.

\section{Discussion}
\label{sec:discussion}

\textbf{Data quality review.} Any data analysis process depends on the quality of the data itself. In this study, we have combined several data sources to build our dataset. From the GWP data, there is a bias introduced by the inaccuracy of the questionnaire which cannot be overlooked. From the weather data, various kinds of indicators can measure shocks. In this study, we use SPEI values that represent an aggregation of several indicators. However, one can also benefit from using raw indicators such as temperature and precipitation. These indicators can bring a more disaggregated viewpoint to the study. Finally, the equal-bin discretization has the advantage of restricting the sensitivity of our models with varying SPEI values. However, this has the disadvantage of giving the same importance to different ranges of shocks. 

\textbf{Methodology review.} Concerning the methodology used, the use of machine learning (ML) brings a new perspective in applied science dealing with several issues such as large datasets, nonlinearity, and multicollinearity. A clear distinction is necessary between the prediction results obtained by these approaches and the causal inference allowed in parameter estimation. To circumvent the limitation of ML approaches to provide causal inference, we have used feature importance and partial dependence metrics to interpret the results obtained. However, a further effort is needed in ML to build alternative methods integrating causal inference advocated by several applied science authors \citep{atheyml19,athey2019generalized,athey2018,athey2015machine}. In the plethora of ML methods, in this study, we have focused on tree-based methods since they suit our study. However, future work may investigate other methods such as neural networks and unsupervised learning approaches to improve performance.

\textbf{Results review.} The results show that the weather feature adds more prediction power than only using the GWP dataset based on the XGB algorithm with higher AUC measures. In general, a longer time horizon of SPEIs (e.g., 18, 24 months) drives more international moving intention while shorter timescale SPEIs (e.g., 1, 12 months) affected the internal or general move intention. It is reasonable since it is likely that international migration decision takes longer time to collect and assess information. As the social science literature illustrates, we also find that it is vital to investigate based on each country since there are unique features with different intensity affecting the migration intention. Moreover, men have a higher tendency to move while the age group of 35-49 shows a higher tendency to stay \citep{beine2018meta}. The younger group (age 15-24) shows a higher tendency to move out internationally, as found in the literature \citep{mayda2010}.

Unfortunately, there were a few consecutive periods of longer than five months with the SPEI's feature importance values above the average. Most of the periods found were three consecutive months followed by two occurrences of four consecutive months and one occurrence of five consecutive months. Overall, it is difficult to draw any global patterns which conform to the inconclusive findings from previous literature.

\section{Conclusion}
\label{sec:conclusion}

In summary, weather features influence the prediction performance on migration intentions. We examine three tree-based machine learning algorithms and derive results with the better performing XGB algorithm using the GWP dataset and weather shock dataset based on various SPEI timescales and lags. The weather indicators show a positive impact on the prediction performance that is significantly higher than without the weather shock dataset. In addition, the longer timescales of SPEI (e.g., 18, 24 months) drive more international migration intention while shorter timescales of SPEI (e.g., 1, 12 months) affected the general move intention. Yet, it is not easy to generalize global weather patterns for the six countries we investigated. Moreover, country-specific models are necessary due to distinct features with different effects on migration intentions. Overall, among the individual characteristics, gender, age, and networks abroad reveal as important features. For example, male shows higher migration intention while the group of age 15-24 shows a higher intention to international migration. 

There can be further improvements with a different way of preprocessing the features, involving more raw indicators besides the SPEI values, using neural networks, and integrating causal inference with ML methods.

\section*{Acknowledgement}
This research is supported by the ARC convention on ``New approaches to understanding and modeling global migration trends'' (convention 18/23-091). This project has received funding from the European Union’s Horizon 2020 research and innovation program under the Marie Skłodowska-Curie grant agreement No. 754412. We thank Professor Frédéric Docquier for the reviews, suggestions, and earlier discussions of the project. We would like to thank Dr. Shari De Baets for her feedback and comments.

\balance
\bibliography{references}  

\begin{thebibliography}{36}
\providecommand{\natexlab}[1]{#1}
\providecommand{\url}[1]{\texttt{#1}}
\expandafter\ifx\csname urlstyle\endcsname\relax
  \providecommand{\doi}[1]{doi: #1}\else
  \providecommand{\doi}{doi: \begingroup \urlstyle{rm}\Url}\fi

\bibitem[Athey(2015)]{athey2015machine}
S.~Athey.
\newblock Machine learning and causal inference for policy evaluation.
\newblock In \emph{Proceedings of the 21th ACM SIGKDD international conference
  on knowledge discovery and data mining}, pages 5--6, 2015.

\bibitem[Athey(2018)]{athey2018}
S.~Athey.
\newblock \emph{The Impact of Machine Learning on Economics}, pages 507--547.
\newblock University of Chicago Press, January 2018.

\bibitem[Athey and Imbens(2019)]{atheyml19}
S.~Athey and G.~W. Imbens.
\newblock Machine learning methods that economists should know about.
\newblock \emph{Annual Review of Economics}, 11\penalty0 (1):\penalty0
  685--725, 2019.

\bibitem[Athey et~al.(2019)Athey, Tibshirani, Wager,
  et~al.]{athey2019generalized}
S.~Athey, J.~Tibshirani, S.~Wager, et~al.
\newblock Generalized random forests.
\newblock \emph{The Annals of Statistics}, 47\penalty0 (2):\penalty0
  1148--1178, 2019.

\bibitem[Beine and Jeusette(2018)]{beine2018meta}
M.~A. Beine and L.~Jeusette.
\newblock A meta-analysis of the literature on climate change and migration.
\newblock CESifo Working Paper Series 7417, CESifo Group Munich, 2018.

\bibitem[Berlemann and Steinhardt(2017)]{berlemann2017climate}
M.~Berlemann and M.~F. Steinhardt.
\newblock Climate change, natural disasters, and migration—a survey of the
  empirical evidence.
\newblock \emph{CESifo Economic Studies}, 63\penalty0 (4):\penalty0 353--385,
  2017.

\bibitem[Bertoli et~al.(2020)Bertoli, Docquier, Rapoport, and
  Ruyssen]{wthrShock2020}
S.~Bertoli, F.~Docquier, H.~Rapoport, and I.~Ruyssen.
\newblock {Weather Shocks and Migration Intentions in Western Africa: Insights
  from a Multilevel Analysis}.
\newblock CESifo Working Paper Series 8064, CESifo Group Munich, 2020.

\bibitem[Black et~al.(2013)Black, Arnell, Adger, Thomas, and Geddes]{Black2013}
R.~Black, N.~W. Arnell, W.~N. Adger, D.~Thomas, and A.~Geddes.
\newblock Migration, immobility and displacement outcomes following extreme
  events.
\newblock \emph{Environmental Science \& Policy}, 27:\penalty0 32--43, 2013.

\bibitem[Branco et~al.(2016)Branco, Torgo, and Ribeiro]{Branco2016}
P.~Branco, L.~Torgo, and R.~P. Ribeiro.
\newblock A survey of predictive modeling on imbalanced domains.
\newblock \emph{ACM Comput. Surv.}, 49\penalty0 (2), Aug. 2016.
\newblock ISSN 0360-0300.

\bibitem[Breiman(2001)]{Breiman2001}
L.~Breiman.
\newblock Random forests.
\newblock \emph{Machine Learning}, 45\penalty0 (1):\penalty0 5–32, Oct. 2001.
\newblock ISSN 0885-6125.

\bibitem[Breiman et~al.(1984)Breiman, Friedman, Stone, and
  Olshen]{breiman1984classification}
L.~Breiman, J.~Friedman, C.~Stone, and R.~Olshen.
\newblock \emph{Classification and Regression Trees}.
\newblock The Wadsworth and Brooks-Cole statistics-probability series. Taylor
  \& Francis, 1984.
\newblock ISBN 9780412048418.

\bibitem[Cattaneo and Peri(2016)]{cattaneo2016migration}
C.~Cattaneo and G.~Peri.
\newblock The migration response to increasing temperatures.
\newblock \emph{Journal of Development Economics}, 122:\penalty0 127--146,
  2016.

\bibitem[Cattaneo et~al.(2019)Cattaneo, Beine, Fr{\"o}hlich, Kniveton,
  Martinez-Zarzoso, Mastrorillo, Millock, Piguet, and
  Schraven]{cattaneo2019human}
C.~Cattaneo, M.~Beine, C.~J. Fr{\"o}hlich, D.~Kniveton, I.~Martinez-Zarzoso,
  M.~Mastrorillo, K.~Millock, E.~Piguet, and B.~Schraven.
\newblock Human migration in the era of climate change.
\newblock \emph{Review of Environmental Economics and Policy}, 13\penalty0
  (2):\penalty0 189--206, 2019.

\bibitem[Chen and Guestrin(2016)]{ChenGuestrin2016}
T.~Chen and C.~Guestrin.
\newblock Xgboost: {A} scalable tree boosting system.
\newblock In B.~Krishnapuram, M.~Shah, A.~J. Smola, C.~C. Aggarwal, D.~Shen,
  and R.~Rastogi, editors, \emph{Proceedings of the 22nd {ACM} {SIGKDD}
  International Conference on Knowledge Discovery and Data Mining, San
  Francisco, CA, USA, August 13-17, 2016}, pages 785--794. {ACM}, 2016.

\bibitem[Dell et~al.(2014)Dell, Jones, and Olken]{dell2014we}
M.~Dell, B.~F. Jones, and B.~A. Olken.
\newblock What do we learn from the weather? the new climate-economy
  literature.
\newblock \emph{Journal of Economic Literature}, 52\penalty0 (3):\penalty0
  740--98, 2014.

\bibitem[Djamba(2003)]{Djamba2003}
Y.~K. Djamba.
\newblock Gender differences in motivations and intentions to move: Ethiopia
  and south africa compared.
\newblock \emph{Genus}, 59\penalty0 (2):\penalty0 93--111, 2003.

\bibitem[Duan et~al.(2014)Duan, Street, Liu, Xu, and Wu]{duan2014selecting}
L.~Duan, W.~N. Street, Y.~Liu, S.~Xu, and B.~Wu.
\newblock Selecting the right correlation measure for binary data.
\newblock \emph{ACM Transactions on Knowledge Discovery from Data (TKDD)},
  9\penalty0 (2):\penalty0 1--28, 2014.

\bibitem[Eslamian(2014)]{eslamian2014handbook}
S.~Eslamian.
\newblock \emph{Handbook of engineering hydrology: environmental hydrology and
  water management}.
\newblock CRC press, 2014.

\bibitem[Fawcett(2006)]{fawcett2006introduction}
T.~Fawcett.
\newblock An introduction to roc analysis.
\newblock \emph{Pattern recognition letters}, 27\penalty0 (8):\penalty0
  861--874, 2006.

\bibitem[{Fisher} et~al.(2018){Fisher}, {Rudin}, and {Dominici}]{fisher18}
A.~{Fisher}, C.~{Rudin}, and F.~{Dominici}.
\newblock {All Models are Wrong, but Many are Useful: Learning a Variable's
  Importance by Studying an Entire Class of Prediction Models Simultaneously}.
\newblock \emph{arXiv e-prints}, art. arXiv:1801.01489, Jan. 2018.

\bibitem[Gallup(2015)]{gallup2015worldwide}
Gallup.
\newblock Worldwide research methodology and codebook, 2015.

\bibitem[Harris et~al.(2020)Harris, Osborn, Jones, and
  Lister]{harris2020version}
I.~Harris, T.~J. Osborn, P.~Jones, and D.~Lister.
\newblock Version 4 of the {CRU TS} monthly high-resolution gridded
  multivariate climate dataset.
\newblock \emph{Scientific data}, 7\penalty0 (1):\penalty0 1--18, 2020.

\bibitem[Hastie et~al.(2009)Hastie, Tibshirani, and
  Friedman]{DBLP:books/lib/HastieTF09}
T.~Hastie, R.~Tibshirani, and J.~H. Friedman.
\newblock \emph{The Elements of Statistical Learning: Data Mining, Inference,
  and Prediction, 2nd Edition}.
\newblock Springer Series in Statistics. Springer, 2009.
\newblock ISBN 9780387848570.

\bibitem[Mayda(2010)]{mayda2010}
A.~Mayda.
\newblock International migration: a panel data analysis of the determinants of
  bilateral flows.
\newblock \emph{J Popul Econ}, 23:\penalty0 1249–1274, 2010.

\bibitem[McFadden(1973)]{mcfadden1973conditional}
D.~McFadden.
\newblock Conditional logit analysis of qualitative choice behaviour.
\newblock In P.~Zarembka, editor, \emph{Frontiers in Econometrics}, pages
  105--142. Academic Press New York, New York, NY, USA, 1973.

\bibitem[Millock(2015)]{millock2015migration}
K.~Millock.
\newblock Migration and environment.
\newblock \emph{Annual Review of Resource Economics}, pages 35--60, 2015.

\bibitem[Mullainathan and Spiess(2017)]{sendhil17}
S.~Mullainathan and J.~Spiess.
\newblock Machine learning: An applied econometric approach.
\newblock \emph{Journal of Economic Perspectives}, 31\penalty0 (2):\penalty0
  87--106, May 2017.

\bibitem[Provost and Fawcett(2013)]{provost2013data}
F.~Provost and T.~Fawcett.
\newblock \emph{Data Science for Business: What you need to know about data
  mining and data-analytic thinking}.
\newblock O'Reilly Media, Inc., 2013.

\bibitem[Quinlan(1986)]{Quinlan1986}
J.~R. Quinlan.
\newblock Induction of decision trees.
\newblock \emph{Machine Learning}, 1:\penalty0 81--106, 1986.

\bibitem[Snoek et~al.(2015)Snoek, Rippel, Swersky, Kiros, Satish, Sundaram,
  Patwary, Prabhat, and Adams]{DBLP:conf/icml/SnoekRSKSSPPA15}
J.~Snoek, O.~Rippel, K.~Swersky, R.~Kiros, N.~Satish, N.~Sundaram, M.~M.~A.
  Patwary, Prabhat, and R.~P. Adams.
\newblock Scalable bayesian optimization using deep neural networks.
\newblock In F.~R. Bach and D.~M. Blei, editors, \emph{Proceedings of the 32nd
  International Conference on Machine Learning, {ICML} 2015, Lille, France,
  6-11 July 2015}, volume~37 of \emph{{JMLR} Workshop and Conference
  Proceedings}, pages 2171--2180, 2015.

\bibitem[StataCorp(2007)]{statacorp2007stata}
StataCorp.
\newblock Stata data analysis and statistical software.
\newblock \emph{Special Edition Release}, 10:\penalty0 733, 2007.

\bibitem[Swets(1988)]{swets1988measuring}
J.~A. Swets.
\newblock Measuring the accuracy of diagnostic systems.
\newblock \emph{Science}, 240\penalty0 (4857):\penalty0 1285--1293, 1988.

\bibitem[Tjaden et~al.(2019)Tjaden, Auer, and Laczko]{Tjaden2019}
J.~Tjaden, D.~Auer, and F.~Laczko.
\newblock Linking migration intentions with flows: Evidence and potential use.
\newblock \emph{International Migration}, 57\penalty0 (1):\penalty0 36--57,
  2019.

\bibitem[Vicente-Serrano et~al.(2010{\natexlab{a}})Vicente-Serrano,
  Beguer{\'\i}a, L{\'o}pez-Moreno, Angulo, and El~Kenawy]{vicente2010new}
S.~M. Vicente-Serrano, S.~Beguer{\'\i}a, J.~I. L{\'o}pez-Moreno, M.~Angulo, and
  A.~El~Kenawy.
\newblock A new global 0.5 gridded dataset (1901--2006) of a multiscalar
  drought index: comparison with current drought index datasets based on the
  palmer drought severity index.
\newblock \emph{Journal of Hydrometeorology}, 11\penalty0 (4):\penalty0
  1033--1043, 2010{\natexlab{a}}.

\bibitem[Vicente-Serrano et~al.(2010{\natexlab{b}})Vicente-Serrano, Beguería,
  and López-Moreno]{VicenteSerrano2010}
S.~M. Vicente-Serrano, S.~Beguería, and J.~I. López-Moreno.
\newblock A multiscalar drought index sensitive to global warming: The
  standardized precipitation evapotranspiration index.
\newblock \emph{Journal of Climate}, 23\penalty0 (7):\penalty0 1696--1718,
  2010{\natexlab{b}}.

\bibitem[Wilhite and Svoboda(2000)]{wilhite2000drought}
D.~A. Wilhite and M.~D. Svoboda.
\newblock Drought early warning systems in the context of drought preparedness
  and mitigation.
\newblock \emph{Early warning systems for drought preparedness and drought
  management}, pages 1--21, 2000.

\end{thebibliography}

\cleardoublepage \let\cleardoublepage\clearpage
\appendix
\begin{appendices}

\numberwithin{table}{section}
\setcounter{table}{0}
\renewcommand{\thetable}{\Alph{section}.\arabic{table}}

\numberwithin{figure}{section}
\setcounter{figure}{0}    
\renewcommand{\thefigure}{\Alph{section}.\arabic{figure}}
\FloatBarrier

\section{Machine learning approaches}
\begin{table}%
\centering
    \begin{tabular}{cccccc}
    \toprule
     instance & age & hhsize & mabr & drought & move \\
    \midrule
       1 & young & large & yes & harsh & Yes \\
       2 & young & large & no & harsh & Yes \\
       3& middle & large & yes & harsh & Yes \\
       4& old & medium & yes & harsh & No \\
       5& old & small & yes & soft & No \\
       6& old & small & no & soft & Yes \\
       7& middle & small & no & soft & No \\
       8& young & medium & yes & harsh & Yes \\
       9& young & small & yes & soft & No \\
       10 & old & medium & yes & soft & No \\
       11 & young & medium & no & soft & No \\
       12 & middle & medium & no & harsh & No \\
       13 & middle & large & yes & soft & No \\
       14 & old & medium & no & harsh & Yes \\
    \bottomrule
    \end{tabular}
    \caption{A sample dataset with individual characteristics, drought index, and migration intention. `hhsize': household size. `mabr': human network abroad. (Note: The table is for an explanation purpose, not the dataset we used.)}
\label{apxtab:smallexample}
\end{table}

\begin{figure}
    \centering
    \resizebox{0.75\columnwidth}{!}{
    \begin{tikzpicture}[level distance=3cm,
  level 1/.style={sibling distance=4cm},
  level 2/.style={sibling distance=2.2cm},
  myleaf/.style={text=blue, draw, rectangle},
  mylabel/.style={text=orange}]
  \node (ww){
  \begingroup
\setlength{\tabcolsep}{3pt} %
\renewcommand{\arraystretch}{1.5} %
  \begin{tabular}{cccccccccccccc}
       1 & 2 & 3 & 4 & 5 & 6 & 7 & 8 & 9 & 10 & 11 & 12 & 13 & 14 \\
       \midrule
       Y & Y & Y & N & N & Y & N & Y & N & N & N & N & N & Y \\
       \multicolumn{14}{c}{\textbf{age}}\\
  \end{tabular}
  \endgroup
  }
    child {node {
    \begingroup
\setlength{\tabcolsep}{3pt} %
\renewcommand{\arraystretch}{1.5} %
  \begin{tabular}{ccccc}
       1 & 2 & 8 & 9 & 11 \\
       \midrule
       Y & Y & Y & N & N \\
       \multicolumn{5}{c}{\textbf{drought}}\\
  \end{tabular}
  \endgroup
    }
      child {node[myleaf] {
      \begingroup
\setlength{\tabcolsep}{3pt} %
\renewcommand{\arraystretch}{1.5} %
  \begin{tabular}{ccc}
       1 & 2 & 8  \\
       \midrule
       Y & Y & Y \\
       \multicolumn{2}{c}{\textbf{Yes}}\\
  \end{tabular}
  \endgroup
  } edge from parent node[left, mylabel] {harsh}}
      child {node[myleaf] {
      \begingroup
\setlength{\tabcolsep}{3pt} %
\renewcommand{\arraystretch}{1.5} %
  \begin{tabular}{cc}
       9 & 11  \\
       \midrule
       N & N \\
       \multicolumn{2}{c}{\textbf{No}}\\
  \end{tabular}
  \endgroup
  } edge from parent node[right, mylabel] {soft} } 
  edge from parent node[left, mylabel] {young}
    } %
    child {node[myleaf] {
    \begingroup
\setlength{\tabcolsep}{3pt} %
\renewcommand{\arraystretch}{1.5} %
  \begin{tabular}{ccccc}
       4 & 5 & 6 & 10 & 14 \\
       \midrule
       N & N & Y & N & Y \\
       \multicolumn{5}{c}{\textbf{No}}\\
  \end{tabular}
  \endgroup
    } edge from parent node[right, mylabel] {old} } %
    child {node {
    \begingroup
\setlength{\tabcolsep}{3pt} %
\renewcommand{\arraystretch}{1.5} %
  \begin{tabular}{cccc}
       3 & 7 & 12 & 13  \\
       \midrule
       Y & N & N & N \\
       \multicolumn{4}{c}{\textbf{mabr}}\\
  \end{tabular}
  \endgroup
  } 
      child {node[myleaf] {
      \begingroup
\setlength{\tabcolsep}{3pt} %
\renewcommand{\arraystretch}{1.5} %
  \begin{tabular}{cc}
       3 & 13  \\
       \midrule
       Y & N \\
       \multicolumn{2}{c}{\textbf{Yes}}\\
  \end{tabular}
  \endgroup
  }  edge from parent node[left, mylabel] {yes} }
      child {node[myleaf] {
            \begingroup
\setlength{\tabcolsep}{3pt} %
\renewcommand{\arraystretch}{1.5} %
  \begin{tabular}{cc}
       7 & 12  \\
       \midrule
       N & N \\
       \multicolumn{2}{c}{\textbf{No}}\\
  \end{tabular}
  \endgroup
      } edge from parent node[right, mylabel] {no} }
      edge from parent node[right, mylabel] {middle}
    } %
     
    ;
    \end{tikzpicture}
    }
    \caption{Example of a decision tree trained with the sample dataset in Table~\ref{apxtab:smallexample}. The numbers from 1 to 14 are instance numbers from Table~\ref{apxtab:smallexample}. Capitalized Y and N represent the moving intention for each instance. }
    \label{fig:smalldt}
\end{figure}
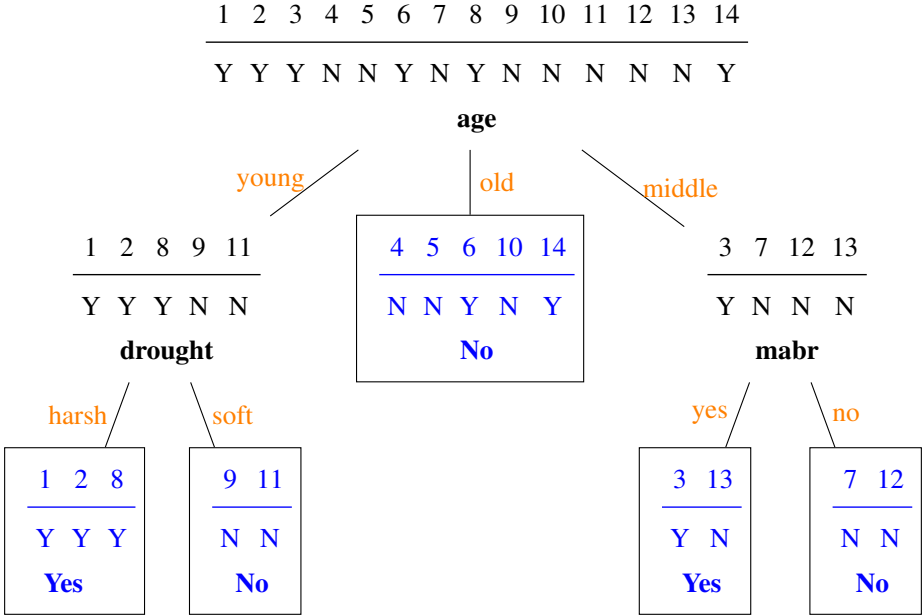

\label{apx:ml}
\label{subsec:methods}

\subsection{Data preprocessing}
We use the sample dataset in Table~\ref{apxtab:smallexample} as an illustrative example. 
This dataset has four features: age, household size (`hhsize'), having human network abroad (`mabr'), and the intensity of the drought (`drought'); and one target attribute representing the migration intention (`move').

The first step is data preprocessing.
It allows cleaning up the data by handling missing data and scale/type-related problems. 
A scale-related issue occurs when variables are displayed in different scales, for example, year (e.g., $[2010, 2016]$) and age (e.g., $[0, 100]$). 
This problem can cause bias on ML models' output and implementation inefficiency. %

There are two types of variables, numerical and categorical variables, that may need preprocessing.
The categorical variables contain labels instead of numerical values. 
Many ML algorithms only support numerical variables, often for the sake of implementation efficiency. 
Hence, it is recommended to convert these variables into numerical variables using \textit{one-hot encoding}. 

\begin{definition}[One-hot encoding]
It consists of creating new binary variables for the unique labels in the categorical variable. 
\end{definition}

It is well-known that it produces bias to the model output when using the numerical variable inputs with different scales.
We overcome this problem by \textit{binarizing} these numerical variables. 

\begin{definition} [Data binarization]
It comprises transforming a numerical variable into several binary variables.
The binarization workflow is in two steps: (i) split the numerical variable into intervals and create a categorical variable by labeling each range. Then, (ii) use the one-hot encoding method to create the binary variables.
\end{definition}

\begin{figure}%
    \centering
    \begin{tikzpicture}[scale=0.4,
    node distance=0pt,
  start chain = A going right,
  arrow/.style = {draw=#1,-{Stealth[bend]}, line width=.4pt, shorten >=1mm, shorten <=1mm}, %
  arrow/.default = black,
  X/.style = {rectangle, draw,%
    minimum width=2ex, minimum height=3ex, outer sep=0pt, on chain},
  B/.style = {decorate,
    decoration={brace, amplitude=5pt, pre=moveto,pre length=1pt,post=moveto,post length=1pt, raise=1mm, #1}, %
    thick},
  B/.default = mirror, %
  ]
    \matrix(mat1)[matrix of nodes,
             nodes={draw, minimum size=15mm,anchor=center, minimum height = 3mm},
             nodes in empty cells={draw=none, minimum size=15mm,anchor=center},
             row 1/.style={nodes={draw=none, text=blue, minimum size=1mm}},
             row 2/.style={nodes={draw=none, text=gray, minimum size=1mm}}] {
             age \\
    numerical \\
    17  \\
    66 \\
    38 \\
    42 \\
    23 \\
    31 \\
    };
    \matrix(mat2)[right= 1cm of mat1,matrix of nodes,
             nodes={draw, minimum size=15mm,anchor=center, minimum height = 3mm},
             nodes in empty cells={draw=none, minimum size=15mm,anchor=center},
             row 1/.style={nodes={draw=none, text=blue, minimum size=1mm}},
             row 2/.style={nodes={draw=none, text=gray, minimum size=1mm}},
             column 1/.style={nodes={draw=none, text=gray}}] {
             & age\\
     & categorical\\
    $[0,30)$ & young \\
    $[65, +\infty)$ & old \\
    $[30,65)$ & middle\\
    $[30,65)$ & middle \\
    $[0,30)$ & young\\
    $[30,65)$ & middle\\
    };
    \matrix(mat3)[right= 1cm of mat2, matrix of nodes,
             nodes={draw, minimum size=15mm, anchor=center, minimum height = 3mm},
             nodes in empty cells={draw=none, minimum size=15mm,anchor=center},
             row 1/.style={nodes={draw=none, text=blue, minimum size=1mm}},
             row 2/.style={nodes={draw=none, text=gray, minimum size=1mm}}] {
             young & middle & old\\
    numerical & numerical & numerical \\
    1 & 0 & 0 \\
     0 & 0 & 1 \\
     0 & 1 & 0 \\
    0 & 1 & 0 \\
     1 & 0 & 0 \\
    0 & 1 & 0 \\
    };
  \draw   (mat1.north -| mat1-1-1.north west) -| (mat1.south west) -- (mat1.south -| mat1-3-1.south west)  (mat1.north -| mat1-1-1.north east) -| (mat1.south east) -- (mat1.south -| mat1-3-1.south east) ;
  \draw   (mat2.north -| mat2-1-2.north west) -| (mat2.south west) -- (mat2.south -| mat2-3-1.south west)  (mat2.north -| mat2-3-2.north east) -| (mat2.south east) -- (mat2.south -| mat2-3-2.south east) ;
  \draw   (mat3.north -| mat3-1-1.north west) -| (mat3.south west) -- (mat3.south -| mat3-3-1.south west)  (mat3.north -| mat3-3-3.north east) -| (mat3.south east) -- (mat3.south -| mat3-3-3.south east) ;
    \end{tikzpicture}
    \caption{Example of one-hot encoding and binarization of the variable age\label{fig:onehot}}
\end{figure}
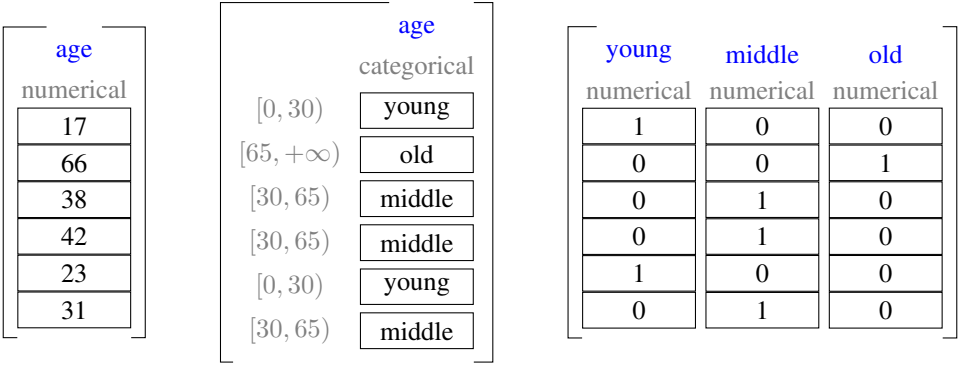

\begin{example}
Figure~\ref{fig:onehot} shows an example of binarization and one-hot encoding for the age variable.
\end{example}

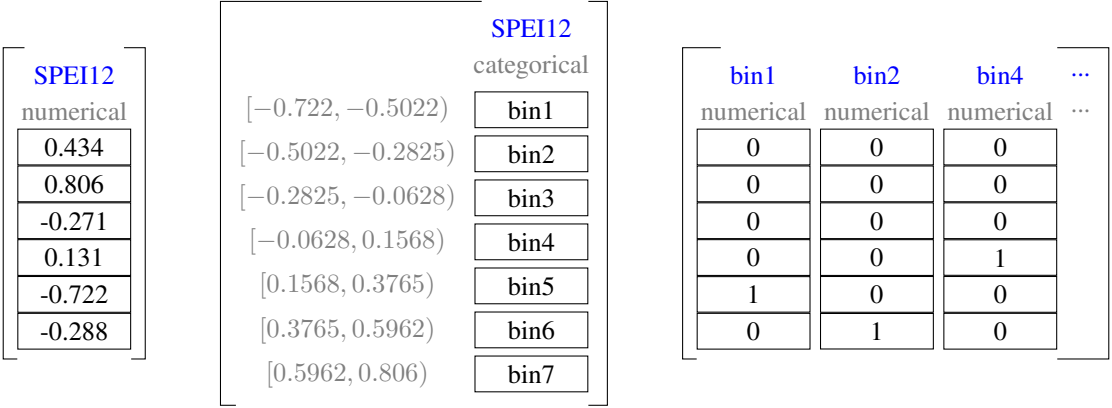
\begin{figure}
    \centering
    \begin{tikzpicture}[scale=0.4,
    node distance=0pt,
  start chain = A going right,
  arrow/.style = {draw=#1,-{Stealth[bend]}, line width=.4pt, shorten >=1mm, shorten <=1mm}, %
  arrow/.default = black,
  X/.style = {rectangle, draw,%
    minimum width=2ex, minimum height=3ex, outer sep=0pt, on chain},
  B/.style = {decorate,
    decoration={brace, amplitude=5pt, pre=moveto,pre length=1pt,post=moveto,post length=1pt, raise=1mm, #1}, %
    thick},
  B/.default = mirror, %
  ]
    \matrix(mat1)[matrix of nodes,
             nodes={draw, minimum size=15mm,anchor=center, minimum height = 3mm},
             nodes in empty cells={draw=none, minimum size=15mm,anchor=center},
             row 1/.style={nodes={draw=none, text=blue, minimum size=1mm}},
             row 2/.style={nodes={draw=none, text=gray, minimum size=1mm}}] {
             SPEI12 \\
    numerical \\
    0.434  \\
    0.806 \\
    -0.271 \\
    0.131 \\
    -0.722 \\
    -0.288 \\
    };
    \matrix(mat2)[right= 1cm of mat1,matrix of nodes,
             nodes={draw, minimum size=15mm,anchor=center, minimum height = 3mm},
             nodes in empty cells={draw=none, minimum size=15mm,anchor=center},
             row 1/.style={nodes={draw=none, text=blue, minimum size=1mm}},
             row 2/.style={nodes={draw=none, text=gray, minimum size=1mm}},
             column 1/.style={nodes={draw=none, text=gray}}] {
             & SPEI12\\
     & categorical\\
    $[-0.722,-0.5022)$ & bin1 \\
    $[-0.5022,-0.2825)$ & bin2 \\
    $[-0.2825,-0.0628)$ & bin3\\
    $[-0.0628,0.1568)$ & bin4 \\
    $[0.1568,0.3765)$ & bin5\\
    $[0.3765,0.5962)$ & bin6\\
    $[0.5962,0.806)$ & bin7\\
    };
    \matrix(mat3)[right= 1cm of mat2, matrix of nodes,
             nodes={draw, minimum size=15mm, anchor=center, minimum height = 3mm},
             nodes in empty cells={draw=none, minimum size=15mm,anchor=center},
             row 1/.style={nodes={draw=none, text=blue, minimum size=1mm}},
             row 2/.style={nodes={draw=none, text=gray, minimum size=1mm}}] {
             bin1 & bin2 & bin4 & ...\\
    numerical & numerical & numerical & ... \\
    0 & 0 & 0 \\
     0 & 0 & 0 \\
     0 & 0 & 0 \\
    0 & 0 & 1 \\
     1 & 0 & 0 \\
    0 & 1 & 0 \\
    };
  \draw   (mat1.north -| mat1-1-1.north west) -| (mat1.south west) -- (mat1.south -| mat1-3-1.south west)  (mat1.north -| mat1-1-1.north east) -| (mat1.south east) -- (mat1.south -| mat1-3-1.south east) ;
  \draw   (mat2.north -| mat2-1-2.north west) -| (mat2.south west) -- (mat2.south -| mat2-3-1.south west)  (mat2.north -| mat2-3-2.north east) -| (mat2.south east) -- (mat2.south -| mat2-3-2.south east) ;
  \draw   (mat3.north -| mat3-1-1.north west) -| (mat3.south west) -- (mat3.south -| mat3-3-1.south west)  (mat3.north -| mat3-3-3.north east) -| (mat3.south east) -- (mat3.south -| mat3-3-3.south east) ;
    \end{tikzpicture}
    \caption{Example of discretization of 12 months timescale of SPEI}
    \label{apxfig:speiDiscretization}
\end{figure}

\textit{Generalization} is an essential concept in ML. It refers to the ability of a method to classify unknown examples to the model correctly.
For this, the dataset is split for training and testing the model in the data preprocessing step.

\begin{definition}[Training set and test set]
 The \textit{training set} is a part of the dataset used to train the model and the \textit{test set} is the hold-out part of the dataset to test the model. Typically, 60 to 90\% of the database is assigned as a \textit{training set} while the rest as a \textit{test set}.
\end{definition}

To have a noise-free and robust model that generalizes well, the training and test sets are extracted iteratively from the dataset. This resampling procedure is called the \textit{cross-validation} process.

\begin{definition}[Cross-Validation]
The cross-validation process consists of randomly splitting the dataset into $K$ fairly equal samples $S_1, S_2, \cdots, S_K$.
Based on these samples, $K$ folds are created, each containing training and testing sets.
At the ith fold, the samples $S_1, S_2, \cdots, S_K$, excluding $S_i$, are merged to a training set and sample $S_i$ is used as a testing set. 
\end{definition}

\begin{figure}
    \centering
    \resizebox{0.6\columnwidth}{!}{
    \begin{tikzpicture}
    \matrix[matrix of nodes, ampersand replacement=\&, 
             nodes={draw, minimum size=8mm,anchor=center},
             nodes in empty cells, minimum height = 1cm,
             row 1/.style={nodes={draw=none}},
             row 2/.style = {nodes={text=blue}}] {
     1 \& 2 \& 3 \& 4 \& 5 \& 6 \& 7 \& 8 \& 9 \& 10 \\
    train \& \textcolor{red}{test} \& train \& train \& train \& train \& train \& train \& train \& train\\
    };
    \end{tikzpicture}
    }
    \caption{10-folds cross validation\label{fig:cross-val}}
    
\end{figure}
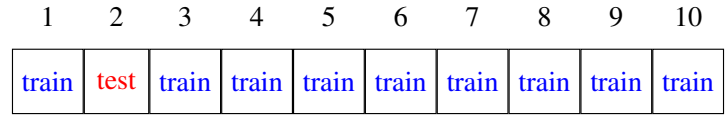

\begin{example}
Figure~\ref{fig:cross-val} shows an example of the second fold.
\end{example}

\subsection{Tree-based approaches}
\label{apx:treeApproach}
Decision tree method approximates the learning function $f$ using decision trees. 

\begin{definition}
 A decision tree represents a set of conditions that satisfies the instance classification. The paths from the root to the leaf represent classification rules. 
\end{definition}

\begin{example}
Figure~\ref{fig:smalldt} is an example of a decision tree using the sample dataset. 
\end{example}

Decision tree algorithms classify instances from the root to the leaves by providing a classification to each instance in leaves. 
Each node represents a test on the features, and each branch corresponds to a potential value of a feature.  

\begin{example}
In the tree in Figure~\ref{fig:smalldt}, age is the root node. 
This node has three branches (young, middle, and old) representing the age values.
The first leaf on at the leftmost of the tree represents all instances where individuals are \textit{young}, and the drought is \textit{harsh}, where people have a moving intention (\textit{move}).
\end{example}

A decision tree is built by selecting the variable at each node that gives the best data split. This split is based on the measure of the impurity rate (obtained by calculating, for example, the entropy or the Gini index) of each variable. The best variable is the one with the lowest impurity rate. Typically, this measure favors splits that allow having the dominant or (strongly) discriminative label over the target attribute. 

It is possible to represent a decision tree as a linear function~\citep{sendhil17}. 
This is closer to the way social scientists represent a model. 
To do so, we represent each leaf of the tree as a variable (feature) of the linear model. 
This variable is the product of decisions from the root to the leaf. 
This model thus contains as many variables as there are leaves in the tree. 
These variables show how decision trees take into account the nonlinearity of the problem automatically.

\begin{example}
Let $L_1, L_2,\cdots,L_5$ be the variables of the linear model. These variables represent the leaves of the tree in Figure~\ref{fig:smalldt} (from left to the right of the tree). 
The leftmost leaf variable $L_1$ is equal to $L_1 = 1_{\text{age = young}\wedge \text{drought = harsh}}$. The variables  $L_3$ and $L_5$ are equal to $L_3 = 1_{\text{age = old}}$, and $L_5 = 1_{\text{age = middle} \wedge \text{mabr = no}}$. Accordingly, the outcome ($y$) follows: 
\begin{equation}
y = f(L) = \beta_1L_1 + \beta_2L_2 + \beta_3L_3 + \beta_4L_4 + \beta_5L_5 + \epsilon
\end{equation}
\end{example}

As in the example, building and using decision trees (DT) are straightforward and explainable. However, in practice, they might be inaccurate~\citep{DBLP:books/lib/HastieTF09}.
Thus, several other tree-based methods have been proposed. Random Forest (RF)~\citep{Breiman2001} and eXtreme Gradient Boosting (XGB)~\citep{ChenGuestrin2016} methods are the well known and widely used ones.

\begin{definition}[Random Forest]
\textit{Random forest} consists of several decision trees that operate together as an ensemble. This ensemble of trees is called a forest. 
Each tree classifies an instance in the forest, and the class label of this instance is decided by a majority vote. Each tree is built on a randomly selected (with replacement) sample of the dataset and a random number of features.
\end{definition}

\begin{example}
With the example of DT in Figure~\ref{fig:smalldt}, instance 1 from our sample example is classified as the class label number (i.e., the individual with instance number 1 has an intention to move). 
With RF that contains five trees, we classify this instance with each tree and take the majority class label. Assuming we have these predictions $\{$Yes, Yes, Yes, No, No$\}$, RF classifies this instance as Yes.
\end{example}

Random forest considers the predictions of each tree to have the same weight. By contrast, XGB does not make this assumption, thus, dynamically assigns a certain weight to each tree and instance. At each step of the forest construction, a new tree is added to address errors made by the existing trees.

By constructing decision trees, one may wonder how deep it needs to go to achieve a better classifier. For a forest, how many trees does it need and how many features to be selected? Basically, in ML, these parameters are determined dynamically by trying several sets of parameters. This process is called parameter tuning. In this paper, we used \textit{Bayesian Hyperparameter Optimization} (BHO)~\citep{DBLP:conf/icml/SnoekRSKSSPPA15}.

\begin{definition}[Bayesian Hyperparameter Optimization]
It consists of testing the models on several parameters, associating each set with a probability to obtain the best performance. A Bayesian Model (i.e., probability model) is then used to select the most promising parameters.
\end{definition}

\subsection{Performance evaluation}
\label{apx:perfEval}
In supervised learning, models are evaluated by making one-on-one comparisons between the predicted outcome ($\hat{y}$) and the real outcome ($y$). This is a benefit of ML over parameter estimation, where the estimation typically relies on the assumptions made from the data-generating process to ensure consistency~\citep{sendhil17}.

For a comparison, in ML, we typically build a confusion matrix.

\begin{definition}[confusion matrix]
A confusion matrix is a matrix that compares the predicted values to the ground truth. It contains four values, namely \textit{true positive} (actual observation `Yes' and predicted `Yes'), \textit{false positive} (actual observation `No' but predicted `Yes', false alarm), \textit{true negative} (actual observation `No' and predicted `No'), and \textit{false negative} values (actual observation `Yes' but predicted `No').
\end{definition}

\begin{figure}
    \centering
    \resizebox{0.9\columnwidth}{!}{
    \begin{tikzpicture}%
    \matrix(predict)[matrix of nodes,ampersand replacement=\&,
             nodes={draw, minimum size=10mm, anchor=center, minimum height = 3mm, font = \Large},
             nodes in empty cells={draw=none, minimum size=1mm,anchor=center},
             row 1/.style={nodes={draw=none, text=blue, minimum size=1mm}},
             row 3/.style={nodes={fill={rgb,255:red,3; green,220; blue,128}}},
             row 4/.style={nodes={draw = none, text = orange}},
             column 1/.style={nodes={draw=none, text=gray}}] {
    \& 1 \& 2 \& 3 \& 4 \& 5 \& 6 \& 7 \& 8 \& 9 \& 10 \& 11 \& 12 \& 13 \& 14 \\
    Observed Move \& Yes \& Yes \& Yes \& No \& No \& Yes \& No \& Yes \& No \& No \& No \& No \& No \& Yes \\
    |[fill=white]| Predicted Move \& Yes \& Yes \& Yes \& No \& No \& |[fill=red]| No \& No \& Yes \& No \& No \& No \& No \& |[fill=red]| Yes \& |[fill=red]| No \\
    Yes probability \& 1.0 \& 1.0 \& 0.50 \& 0.40 \& 0.40 \& 0.40 \& 0.0 \& 1.0 \& 0.0 \& 0.40 \& 0.0 \& 0.0 \& 0.50 \& 0.40 \\
    };
    \node[below= 5mm of predict, 
    xshift = -6.5cm
    ](cmat){
    \noindent
\renewcommand\arraystretch{1.5}
\setlength\tabcolsep{0pt}
\large{
\begin{tabular}{c >{\bfseries}r @{\hspace{0.7em}}c @{\hspace{0.4em}}c @{\hspace{0.7em}}l}
  \multirow{10}{*}{\rotatebox{90}{\parbox{5cm}{\bfseries\centering Actual value}}} & 
    & \multicolumn{2}{c}{\bfseries Prediction outcome} & \\
  & & \bfseries Yes & \bfseries No & \bfseries total \\
  & Yes & \MyBox{True}{Positive\\$=4$} & \MyBox{False}{Negative\\$=2$} & P$'=6$ \\[2.4em]
  & No & \MyBox{False}{Positive\\$=1$} & \MyBox{True}{Negative\\$=7$} & N$'=8$ \\
  & total & P$=5$ & N$=9$ & All$=14$ \\
\end{tabular}
}
    };
    \node[right =2mm of cmat](metrics) {
    \Large{
    \begin{tabular}{c|c|c|}
    \toprule
         Metrics & Formula & Values \\
         \midrule
   \MetBox{Precision} &  \OkBox{$\frac{\text{True Positive}}{\text{True Positive}\ +\ \text{False Positive}}$} & \ValBox{$\frac{4}{4+1}=0.80$} \\
    \midrule
    \MetBox{Recall} & \OkBox{$\frac{\text{True Positive}}{\text{True Positive}\ +\ \text{False Negative}}$} & \ValBox{$\frac{4}{4+2}=0.67$} \\
    \midrule
    \MetBox{Accuracy} & \OkBox{$\frac{\text{True Positive}\ +\ \text{True Negative}}{All}$} & \ValBox{$\frac{4+7}{14}=0.78$} \\
         \bottomrule
    \end{tabular}
    }
    };
    \node[below = 8cm of predict]{\includegraphics[scale=0.8]{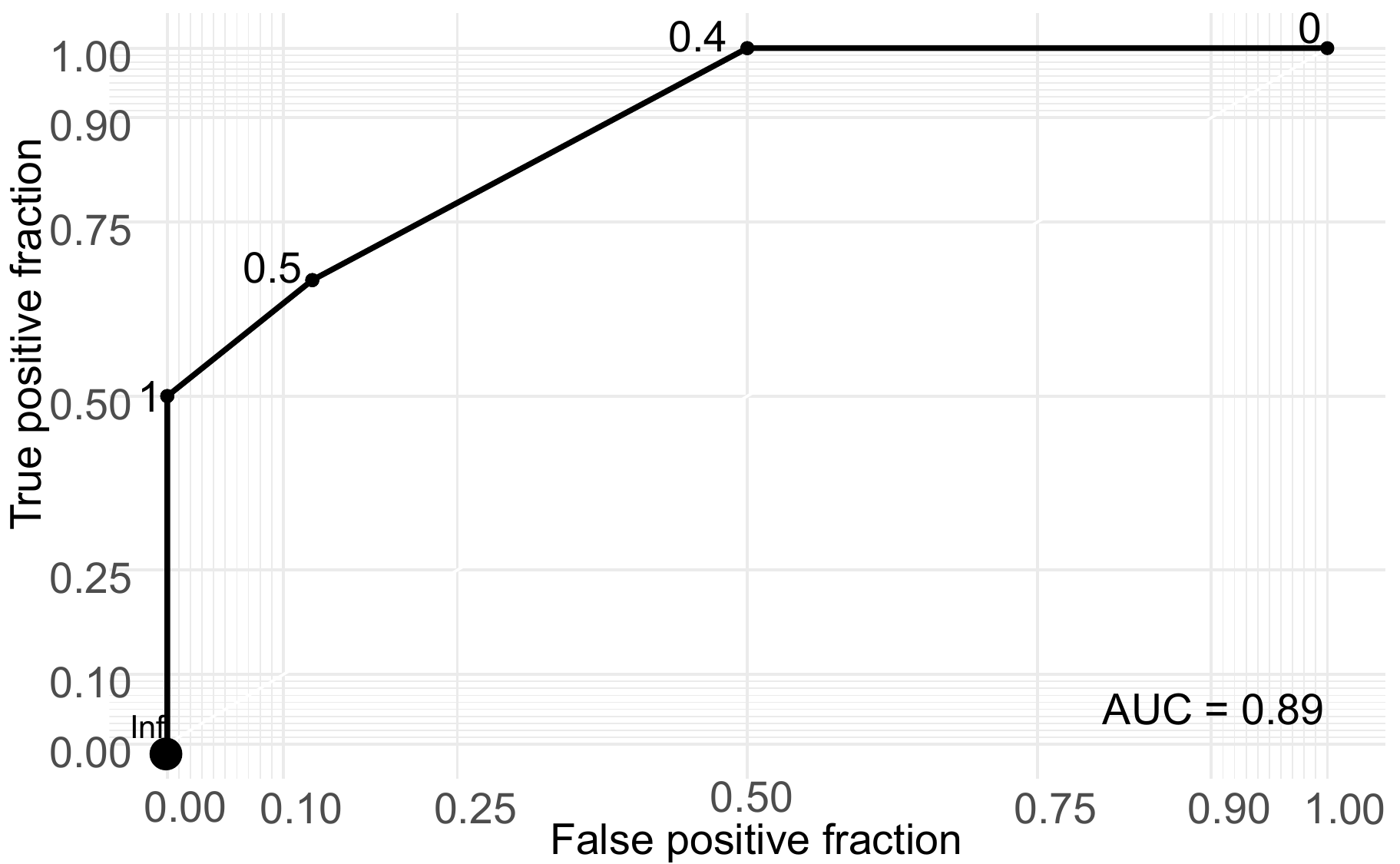}};
    \node[below = 0 of predict]{\Large{Predictions}};
    \node[below = 0 of cmat]{\Large{Confusion Matrix}};
    \node[below = 0 of metrics]{\Large{Metric calculations}};
    \end{tikzpicture}
    }
    \caption{Model performance evaluation with Precision, Recall, Accuracy, and AUC based on the confusion matrix values.\label{fig:confusionmatrixevals}}
\end{figure}
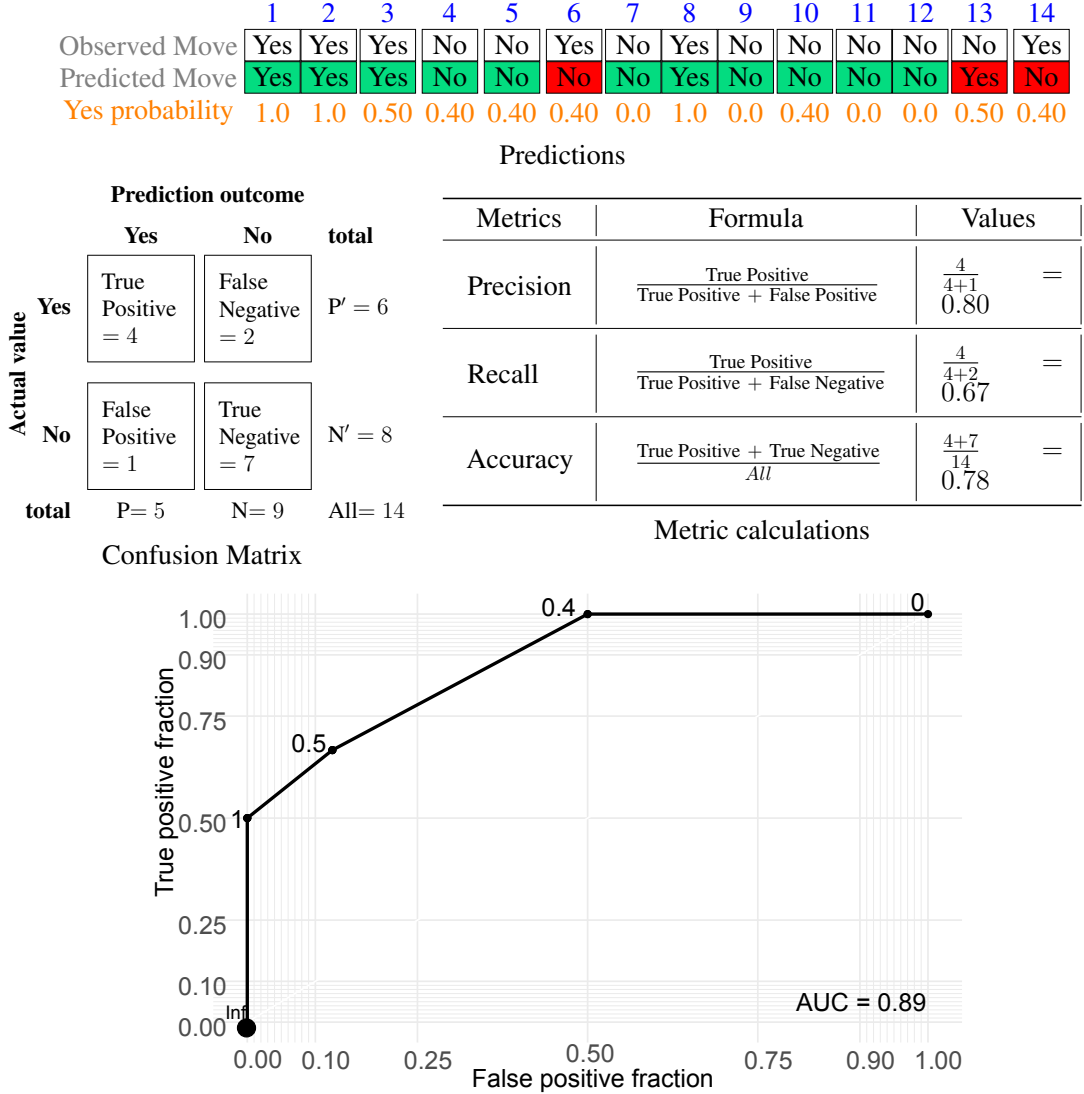

\begin{example}
Figure~\ref{fig:confusionmatrixevals} shows the predicted move intention using decision tree (DT) and the confusion matrix by comparing these predictions to the observed (actual) move intention.
\end{example}

Based on the confusion matrix, various performance metrics can be computed. The common ones are \textit{accuracy}, \textit{precision}, and \textit{recall}.

\begin{definition}[Accuracy - Precision - Recall]
The \textit{accuracy} is a ratio of correctly predicted observations to the total number of observations. It is an intuitive measure, but only when false positive and false negatives are not too different. Instead, \textit{precision} shows the ratio of correctly predicted positive observations to the total predicted positive observations, while \textit{recall} is the ratio of correctly predicted positive observations to all accurate (or true) observations. 
The formulas are available in Figure~\ref{fig:confusionmatrixevals} with the confusion matrix.
These measurements have values between 0 and 1 (the higher, the better performance).
\end{definition}

Predicted class labels typically involve a user-defined threshold (e.g., 0.5). By convention, the probability lesser or equal to the threshold is considered as a 'No' and otherwise a 'Yes'. Differently defined threshold leads to different predictions.
The Area under the ROC (Receiver operating Characteristics) curve (AUC)~\citep{swets1988measuring, fawcett2006introduction}, another model performance metric, is used to evaluate the performance regardless of any classification threshold. 

\begin{definition}[ROC and AUC]
A ROC curve, a two-dimensional graph, is generated by plotting the false-positive fraction (x-axis) against the true-positive fraction (y-axis) of a model for each possible threshold value. The ROC curve shows how well a model classifies binary outcomes.
The AUC (Area under the curve), as its name implies, is the area under the ROC curve. Typically, it is computed when a single value is needed to summarize a model's performance to undertake comparisons. The AUC value is also between 0 and 1 (the higher, the better performance).
\end{definition}

\begin{example}
Figure~\ref{fig:confusionmatrixevals} illustrates the ROC curve and the AUC of a decision tree (DT). The AUC of this classifier is 0.89 (i.e. classifier performs well).
\end{example}

In this paper, we mainly use AUC and precision to determine which method to focus on.

\subsection{Output interpretation: Feature importance and Partial Dependence Plots (PDP)}
\label{apx:pdp}
The features $X$ used to estimate $f$ in the equation $f(X) = y$ are rarely equally relevant. 
Typically, only a small subset of the features is relevant.  
Hence, after training the model, the \textit{Relative Feature Importance (RFI)} method is used to determine the relevant features.  
RFI was introduced by \citet{breiman1984classification} for the tree-based learning methods. 

\begin{definition}[RFI]
RFI consists of, (i) for each internal node of a tree T, compute the contribution of each feature to the prediction,
(ii) then sum its contributions for each feature, and (iii) arrange the features accordingly. 
\end{definition}

To calculate the importance $I_j$ of the feature $j$ (at node $j$) in a decision tree~\eqref{eqn:fi}, five elements are needed: the numbers of `Yes' ($w_j^{Yes}$) and `No' ($w_j^{No}$) instances, the total number of instances ($w_j = w_j^{Yes} + w_j^{No}$) in node $j$, the contribution of $j$ ($c_j = \sum_{i \in \{\text{Yes, No}\}} w_j^{i}$), and the importance of the node $j$ ($n_j = w_j c_j - \sum_{k \in \{\text{children of }j\}}w_k c_k $).
\begin{equation}\label{eqn:fi}
    I_j = \frac{n_j}{\sum_{i \in \{\text{all feature nodes}\}} n_i}
\end{equation}

\begin{example}
The importance of the feature age is 
\begin{equation*}
    I_{\text{age}} = \frac{n_{\text{age}}}{n_{\text{age}}+n_{\text{drought}}+n_{\text{mabr}}} = \frac{-1108}{-1108-18-8} = 0.977
\end{equation*}
\end{example}

In a single decision tree, it is clear that the most important feature is the feature at the root node. 
In a forest, \eqref{eqn:fi} is generalized as follows:
\begin{equation}\label{eqn:rfi}
    RI_j = \frac{\sum_{t \in \{\text{forest}\}} n_j^t }{\sum_{t \in \{\text{forest}\}}\sum_{i \in \{\text{all feature nodes of } t\}} n_i^t}
\end{equation}

\begin{figure}
    \centering
    \resizebox{\columnwidth}{!}{
    \begin{tikzpicture}[level distance=1.5cm,
  level 1/.style={sibling distance=1.2cm},
  level 2/.style={sibling distance=1.2cm}]
  \node (ww){14(6,8)}
    child {node {5(3,2)}
      child {node {3(3,0)}}
      child {node {2(0,2)}}
    } %
    child {node {5(2, 3)}} %
    child {node {4(1,3)}
      child {node {2(1, 1)}}
      child {node {2(0, 2)}}
    } %
    ;
    \node[right = 2.5cm of ww] (cc) {$6\times-5 + 8\times-7 = -86$}
    child {node {-8}
      child {node {-6}}
      child {node {-2}}
    } %
    child {node {-8}} %
    child {node {-4}
      child {node {-2}}
      child {node {-2}}
    } %
    ;
    \node[right = 1cm of cc] (nn) {\shortstack{$14\times-86 -$ \\ $(5\times -8 + 5\times -8 +4\times -4)$ \\ $ = -1108$}}
    child {node {-18}
      child {node {x}}
      child {node {x}}
    } %
    child {node {x}} %
    child {node {-8}
      child {node {x}}
      child {node {x}}
    } %
    ;
    \node[below = 3cm of ww] (lww) {\shortstack{Number of instances:\\ $w_j$ ($w_j^{Yes}$, $w_j^{No}$)}};
    \node[below = 3cm of cc] (lcc) {\shortstack{Nodes' Contribution:\\ $c_j = \sum_{i \in \{\text{Yes, No}\}} w_j^{i}$}};
    \node[below = 2.7cm of nn] (lnn) {\shortstack{Nodes' importance: \\ $n_j = w_j c_j - \sum_{k \in \{\text{children of }j\}}w_k c_k $}};
    \end{tikzpicture}
    }
    \caption{The five elements needed to compute the feature importance in DT in Figure~\ref{fig:smalldt}\label{fig:smallrfi}}
\end{figure}
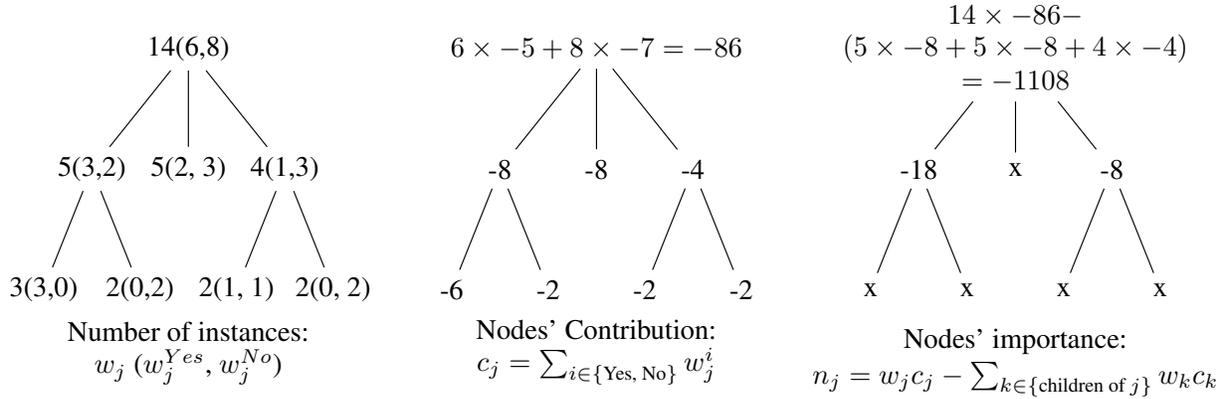

RFI has become widespread and is used for other ML methods. 
In order to understand how these important features influence the outcome $y$, one uses the \textit{Partial Dependency Plots}~\citep[Chap. 14]{DBLP:books/lib/HastieTF09}.

\begin{definition} [Partial Dependence]
Assume the features $X = X_1, X_2,\cdots, X_p$, indexed by $P = \{1, 2,\cdots,p\}$. Let $S$ and its complement $R$ be subsets of $P$, i.e., $S,R \subset P \wedge S\cup R = P \wedge S\cap R = \emptyset$. 
Assuming that $f(X) = f(X_S, X_R)$, the partial dependence of $f(X)$ on the features $X_S$ is,
\begin{equation}\label{eqn:pd}
    PD_S(X_S) = E_{X_R} f(X_S, X_R) \approx \frac{1}{N} \sum^N_{i=1} f(X_S, x_{iR})
\end{equation}
This is a marginal average of $f$ describing the effect of a chosen set of features $S$ on $f$. It is approximated as the average over the $N$ instances in the training set ($X$) of the prediction of each instance ($x_{iR}$) occurring in the complementary set $X_R$.  %
\end{definition}

The computation of \eqref{eqn:pd} requires a pass over the data for each set of joint values of $X_S$.
This can be computationally intensive, and therefore, the partial dependency is usually not calculated with more than three features. 
Fortunately, partial dependence with only one feature is often informative enough, and it simplifies the calculation with a discrete feature. 
In practice, for a discrete feature with two class labels `yes' and `no', we only compute $PD_{S}(X_S = yes)$ and $PD_{S}(X_S = no)$.

\begin{figure}
    \centering
    \resizebox{\columnwidth}{!}{
    \begin{tikzpicture}
    \node (dyes) {
    \begin{tabular}{ccc>{\columncolor{orange}}ccc}
    \toprule
     & age & hhsize & \shortstack{mabr\\=yes} & drought & Move \\
    \midrule
       1 & young & large & yes & harsh & Yes \\
       2 & young & large & yes & harsh & Yes \\
       3& middle & large & yes & harsh & Yes \\
       4& old & medium & yes & harsh & No \\
       5& old & small & yes & soft & No \\
       6& old & small & yes & soft & Yes \\
       7& middle & small & yes & soft & No \\
       8& young & medium & yes & harsh & Yes \\
       9& young & small & yes & soft & No \\
       10 & old & medium & yes & soft & No \\
       11 & young & medium & yes & soft & No \\
       12 & middle & medium & yes & harsh & No \\
       13 & middle & large & yes & soft & No \\
       14 & old & medium & yes & harsh & Yes \\
    \bottomrule
    \end{tabular}
    };
    \node (dyesarrowa) [My Arrow Style, right = 0 of dyes, text width=1.5cm] {Prediction};
    \node (dyespred)[right = 0 of dyesarrowa] {
    \begin{tabular}{c}
    \toprule
     Move \\
    \midrule
       Yes \\
       Yes \\
       Yes \\
       No \\
       No \\
       No \\
       Yes \\
       Yes \\
       No \\
       No \\
       No \\
       Yes \\
       Yes \\
       No \\
    \bottomrule
    \end{tabular}
    };
    \node (dyesarrowb) [My Arrow Style, right = 0 of dyespred, text width=1.5cm] {Proportion};
    \node (dyespdp) [right = 0 of dyesarrowb, font=\Large] {$pd_{\text{mabr} = yes}(X) = 0.50$};
    \node [below =0  of dyes] (dno) {
    \begin{tabular}{ccc>{\columncolor{orange}}ccc}
    \toprule
     & age & hhsize & \shortstack{mabr\\=no} & drought & Move \\
    \midrule
       1 & young & large & no & harsh & Yes \\
       2 & young & large & no & harsh & Yes \\
       3& middle & large & no & harsh & Yes \\
       4& old & medium & no & harsh & No \\
       5& old & small & no & soft & No \\
       6& old & small & no & soft & Yes \\
       7& middle & small & no & soft & No \\
       8& young & medium & no & harsh & Yes \\
       9& young & small & no & soft & No \\
       10 & old & medium & no & soft & No \\
       11 & young & medium & no & soft & No \\
       12 & middle & medium & no & harsh & No \\
       13 & middle & large & no & soft & No \\
       14 & old & medium & no & harsh & Yes \\
    \bottomrule
    \end{tabular}
    };
    \node (dnoarrowa) [My Arrow Style, right = 0 of dno, text width=1.5cm] {Prediction};
    \node (dnopred)[right = 0 of dnoarrowa] {
    \begin{tabular}{c}
    \toprule
     Move \\
    \midrule
       Yes \\
       Yes \\
       No \\
       No \\
       No \\
       No \\
       No \\
       Yes \\
       No \\
       No \\
       No \\
       No \\
       No \\
       No \\
    \bottomrule
    \end{tabular}
    };
    \node (dnoarrowb) [My Arrow Style, right = 0 of dnopred, text width=1.5cm] {Proportion};
    \node (dnopdp) [right = 0 of dnoarrowb, font=\Large] {$pd_{\text{mabr} = no}(X) = 0.21$};
    \end{tikzpicture}
    }
    \caption{Partial Dependence computation for the `mabr' feature (connections abroad)\label{fig:pdpdt}}
\end{figure}
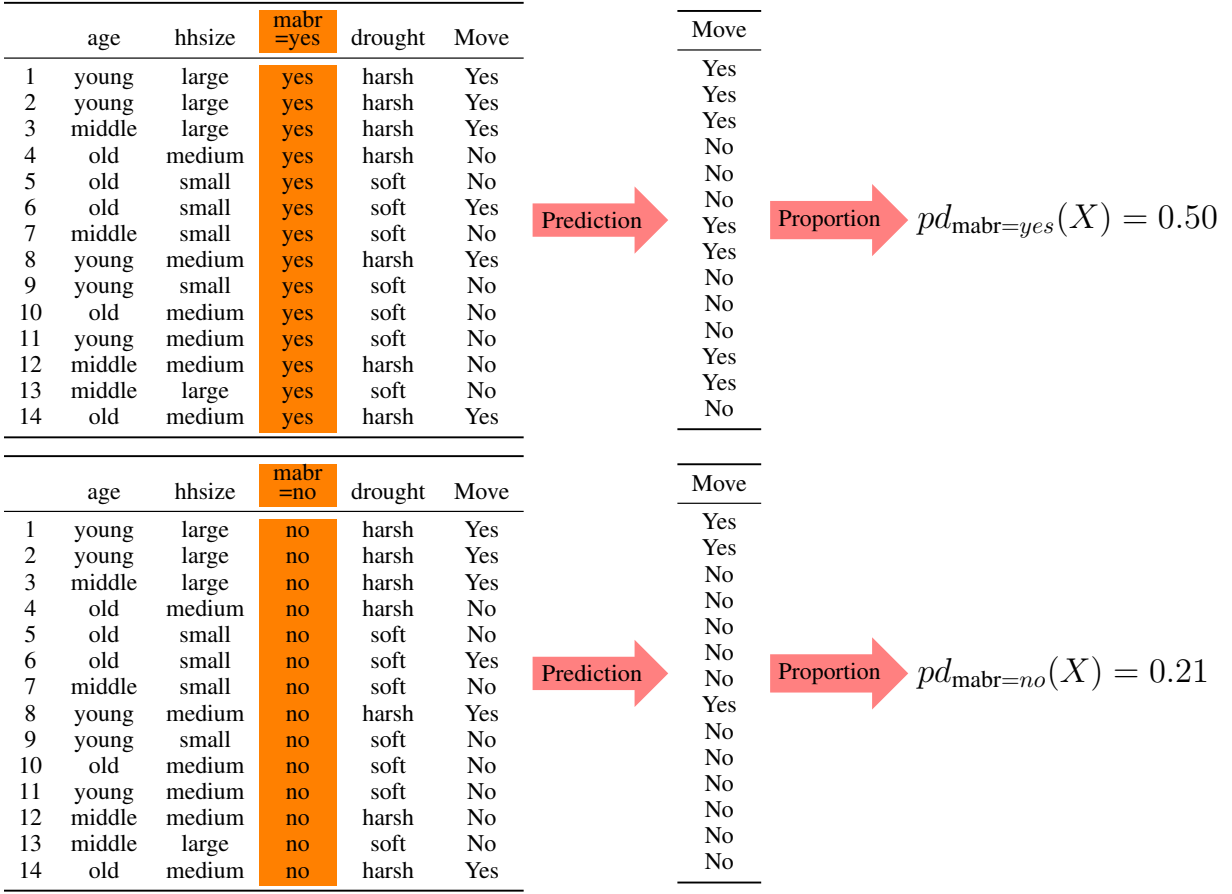

\begin{example}
Figure~\ref{fig:pdpdt} shows how we compute the partial dependence in DT (Figure~\ref{fig:smalldt}) on a feature `mabr' (human network abroad).
\end{example}

From the different values used to calculate the partial dependence, we can draw a chart with the tested values in x-axis against the partial dependence output in y-axis.
The plot's role is to show in which direction (towards label `Yes' or `No') each feature value drives the outcome $y$.
The plot visualizes the effect of a feature related to the average effects of other features.

\newpage
\section{Additional figures}
\label{apx:additional}

\begin{figure*}
   \centering
   \subfloat[international move]
   {
       \includegraphics[width=0.495\textwidth]{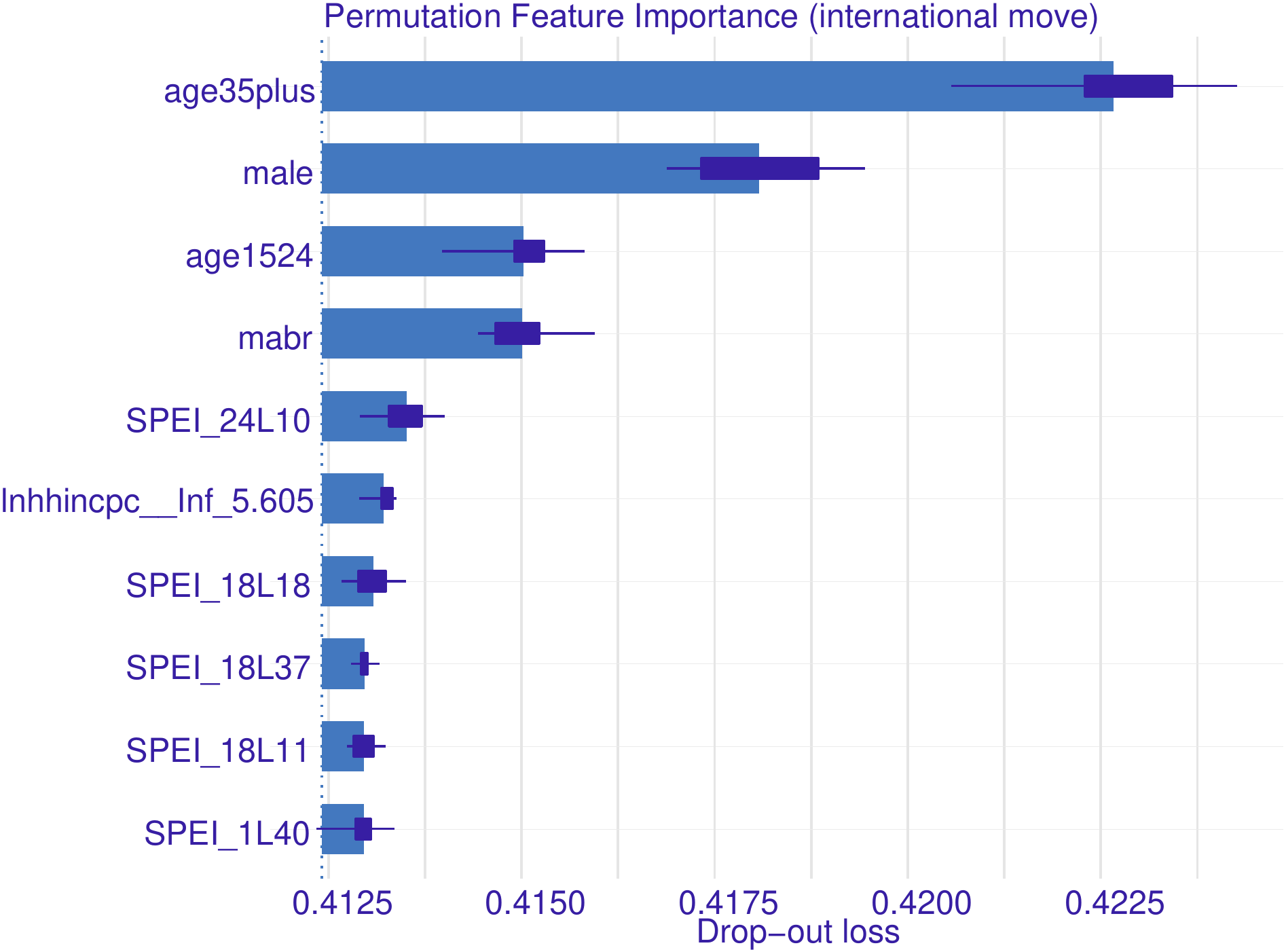}
       \label{pfi1}   
   }
   \subfloat[general move]
   {
       \includegraphics[width=0.495\textwidth]{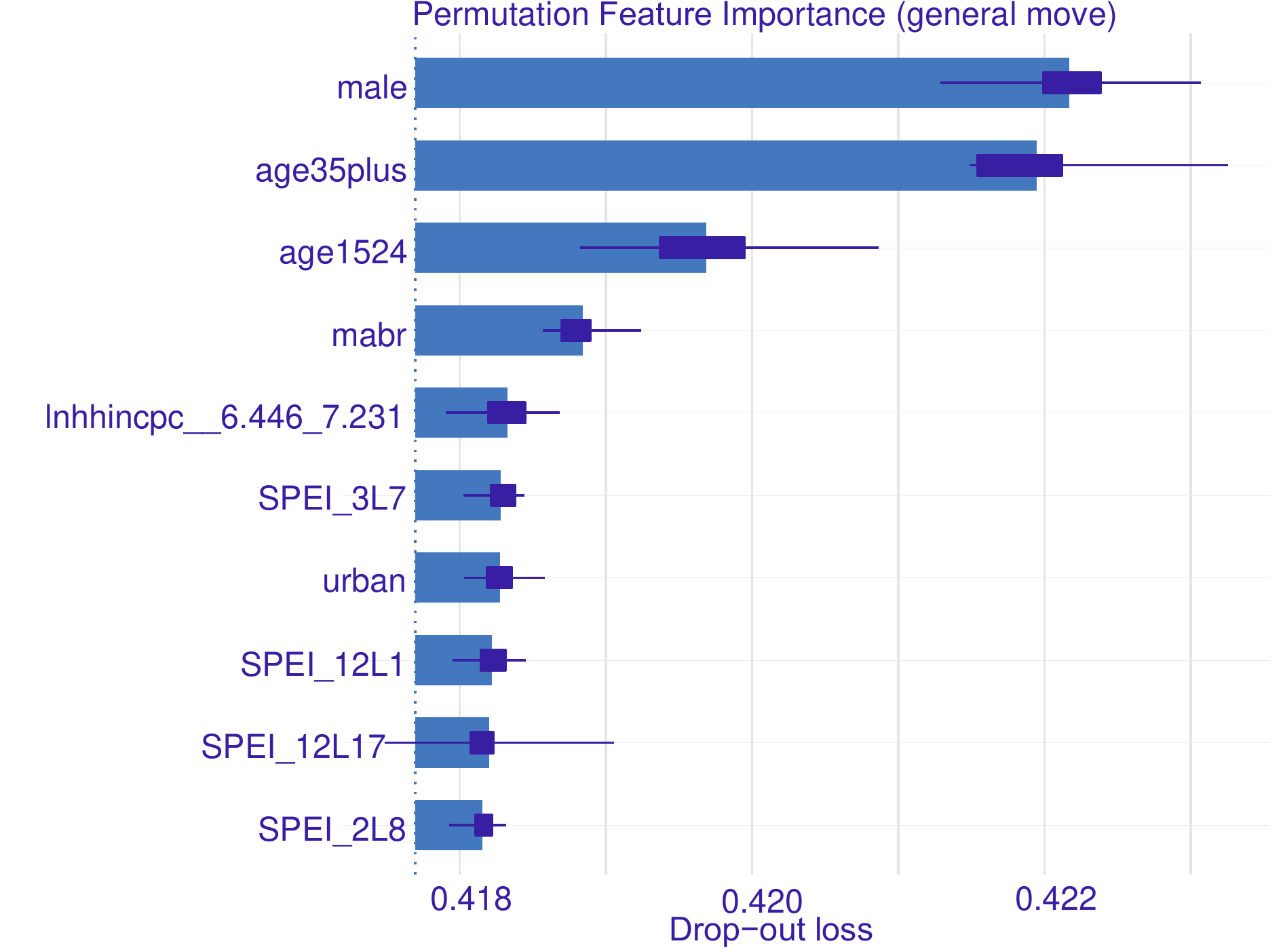}
       \label{pfi2}   
   }
   \caption{Male and age appear as top features based on the permutation feature importance.} 
   \label{apxfig:pfi}   
\end{figure*}

Figure \ref{apxfig:pfi} shows male and age as top influencing features according to the permutation feature importance measures, similar to the results from the Relative Feature Importance (RFI) method. We also observe international move is more affected by longer SPEIs (e.g., 18, 24) while the general move is affected by shorter SPEIs (e.g., 2, 3, 12) which aligns with previous findings.
Darker box plot shows the uncertainty from the permutations.
Permutation feature importance measures the increase of a model's prediction error after a certain feature's value is permuted. The permutation breaks the relationship between the feature and the true outcome. A feature is considered `important' if the change of a feature value increases the model error since it means that the model relies on that feature for prediction. \citet{fisher18} proposed `model reliance' measures and a model-agnostic permutation feature importance algorithm. 

\begin{figure*}
   \centering
   \subfloat[Burkina Faso]
   {
       \includegraphics[width=0.495\textwidth]{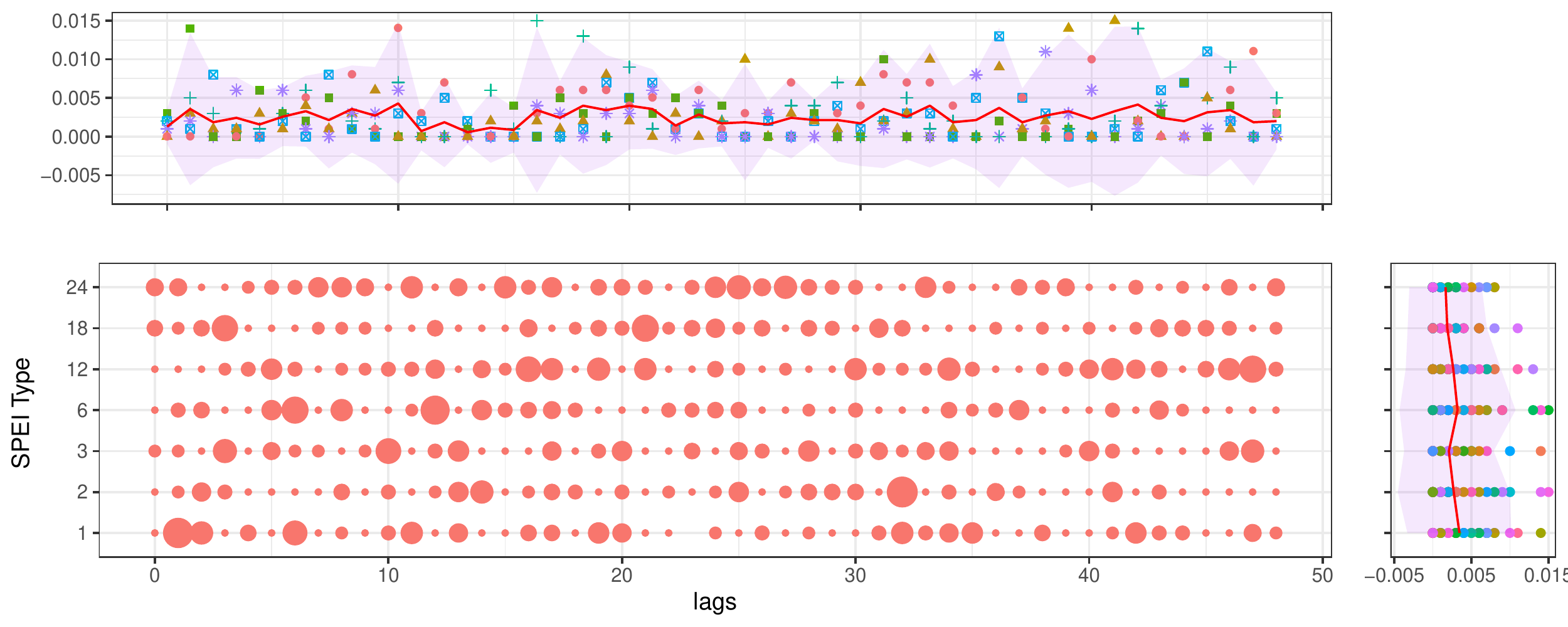}
       \label{sl_bf}   
   }
   \subfloat[Ivory Coast]
   {
       \includegraphics[width=0.495\textwidth]{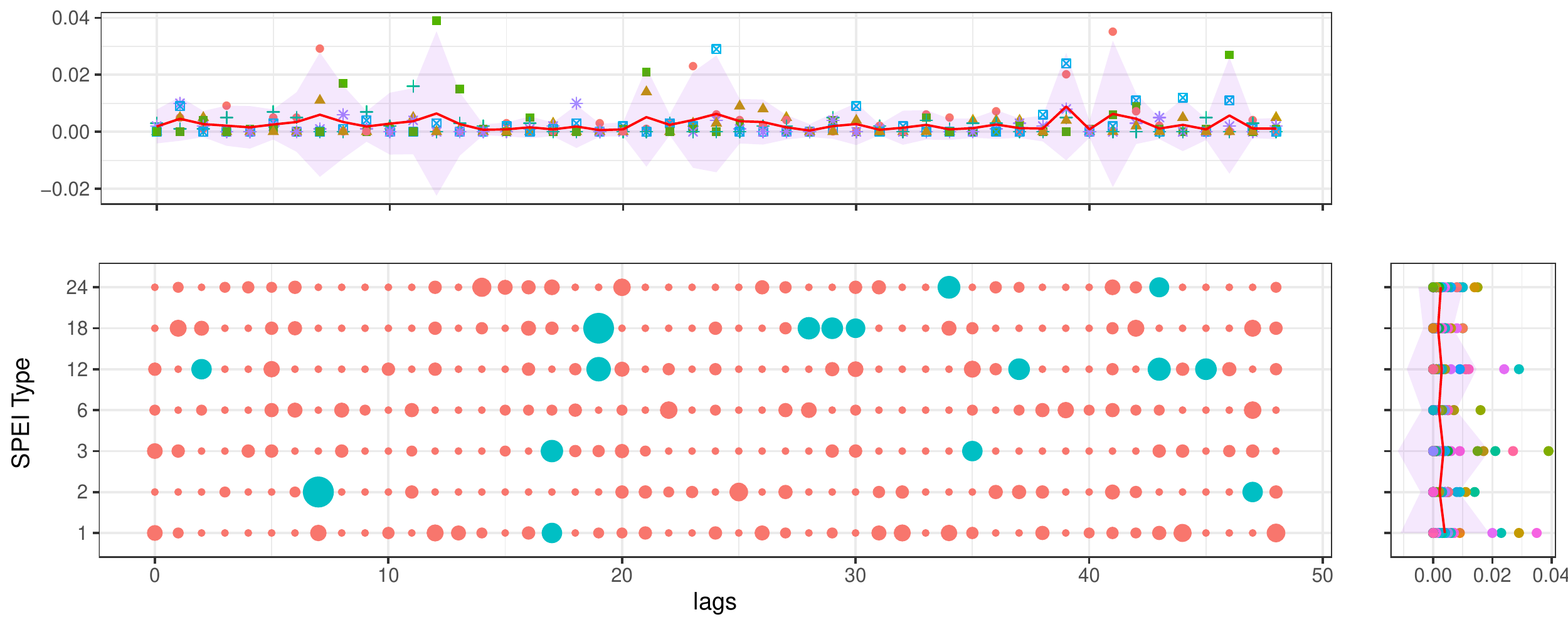}
       \label{sl_ic}   
   }
   
   \subfloat[Mali]
   {
       \includegraphics[width=0.495\textwidth]{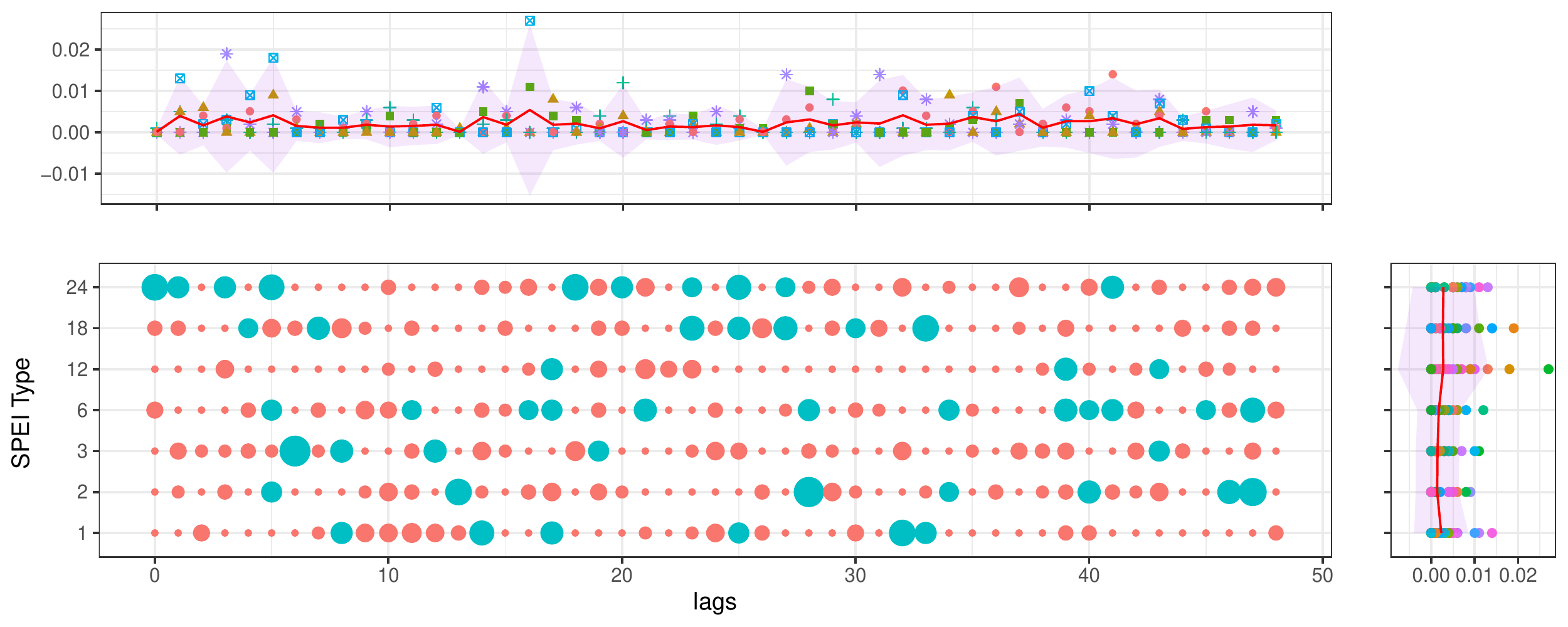}
       \label{sl_mali}   
   }
   \subfloat[Mauritania]   
   {
       \includegraphics[width=0.495\textwidth]{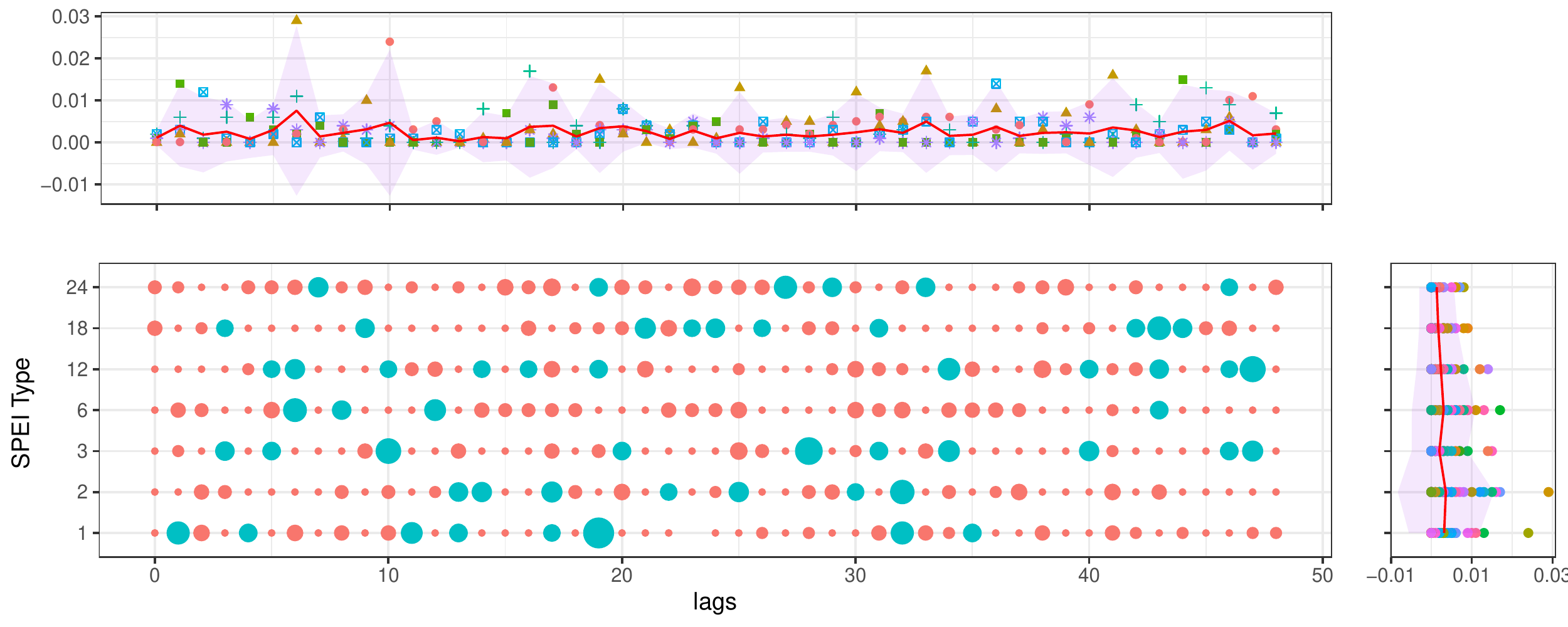}
       \label{sl_mauritania}   
   }     
   
   \subfloat[Niger]
   {
       \includegraphics[width=0.495\textwidth]{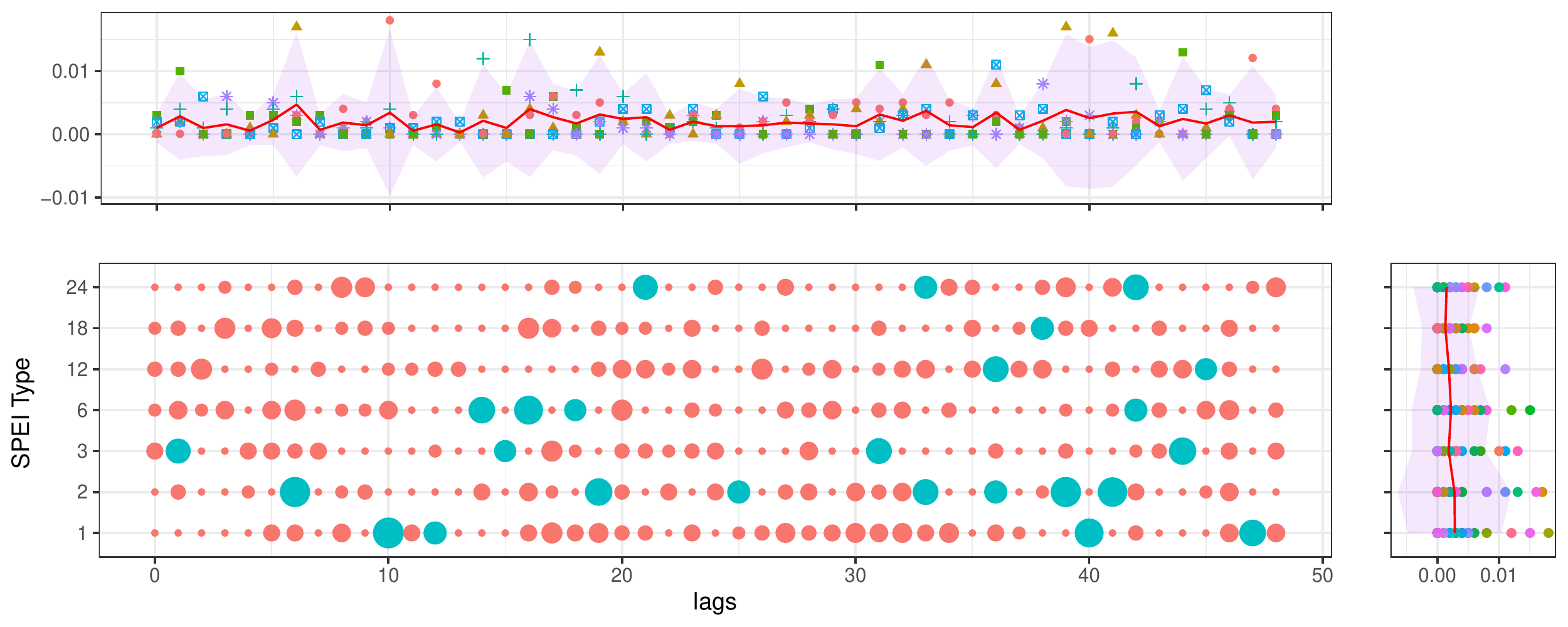}
       \label{sl_niger}   
   }
   \subfloat[Senegal]
   {
       \includegraphics[width=0.495\textwidth]{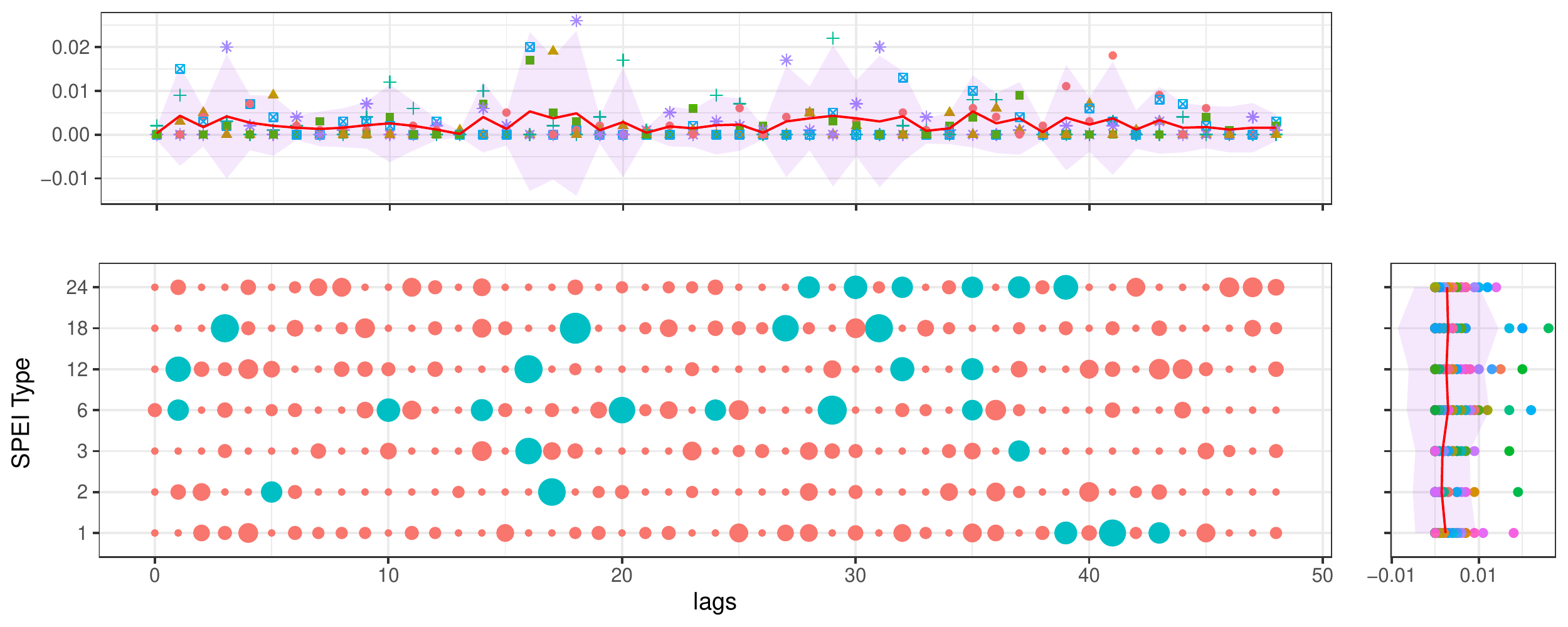}
       \label{sl_senegal}   
   }
   \caption{Feature importance (dot size) based on different SPEI timescales and lags with the distribution of those by each lag (top) and each SPEI (right).} 
   \label{apxfig:speiLag}   
\end{figure*}

Figure \ref{apxfig:speiLag} shows the feature importance distributions of the six countries targeting international move over the seven SPEI timescales (i.e., 1, 2, 3, 6, 12, 18, 24) and 49 lags (i.e., 0-48).

\section{Terminology comparison}
\label{apx:terminology}
Table \ref{apxtab:terminology} compares the common terminology used in social sciences and machine learning.

\renewcommand{\arraystretch}{1.5}
\begin{table}
\centering
\caption{Mapping of terminology used in social science and machine learning}
\label{apxtab:terminology}
\begin{tabular}{c|c}
\toprule
 Social science &  Machine Learning\\
\toprule
\makecell[c]{ independent variable,\\ covariate, \\control variable} &  \makecell[c]{ variable, feature,\\ attribute, column}\\
\hline
observation & \makecell[c]{ observation,\\ row, example,\\ instance}\\
\hline
\makecell[c]{ output, \\dependent variable,\\ outcome} & \makecell[c]{ output,\\ dependent variable, \\target attribute}\\
\hline
sub-sample &  training set \\
\hline
sub-sample & training set \\
\bottomrule
\end{tabular}
\end{table}

\section{GWP questions}
\label{apx:gwpQuestions}

Table \ref{apxtab:gwpQuestions} describes the World Poll questions used to measure opinions from the interviewees.

\begin{table}
\centering
\caption{GWP questions}
\label{apxtab:gwpQuestions}
\begin{tabular}{c|l}
\toprule
\small Feature & \small Description\\
\toprule
age & \small Please tell me your age.\\
\hline
hhsize & \makecell[l]{\small Including yourself, how many people who are residents of this country, age 15 or over, currently  \\\small live in this household?}\\
\hline
hskill & \small Education Category\\
\hline
Inhhincpc & \small Annual household income in International Dollars\\
\hline
mabr & \makecell[l]{\small Do you currently have family members or relatives living permanently in other countries, or not? \\\small(including countries of the former Soviet Union)}\\
\hline
male & \small Gender\\
\hline
urban & \small Do you live in . . . ?\\
\hline
year & \small The year gallup survey was performed. \\
\bottomrule
\end{tabular}
\end{table}

\end{appendices}

\end{document}